\documentclass{emulateapj}

\usepackage{graphicx}
\usepackage{xcolor}
\usepackage{epstopdf}
\usepackage{amsmath}
\usepackage{hyperref}
\usepackage{cleveref}
\usepackage{natbib}
\def\simlt{\mathrel{\hbox{\rlap{\hbox{\lower4pt\hbox{$\sim$}}}\hbox{$<$}}}}
\def\simgt{\mathrel{\hbox{\rlap{\hbox{\lower4pt\hbox{$\sim$}}}\hbox{$>$}}}}

\def\ale{\mathrel{\hbox{\rlap{\hbox{\lower4pt\hbox{$\sim$}}}\hbox{$<$}}}}
\def\age{\mathrel{\hbox{\rlap{\hbox{\lower4pt\hbox{$\sim$}}}\hbox{$>$}}}}

\def\gs{\mathrel{\raise0.35ex\hbox{$\scriptstyle >$}\kern-0.6em
\lower0.40ex\hbox{{$\scriptstyle \sim$}}}}
\def\ls{\mathrel{\raise0.35ex\hbox{$\scriptstyle <$}\kern-0.6em
\lower0.40ex\hbox{{$\scriptstyle \sim$}}}}

\def\spose#1{\hbox to 0pt{#1\hss}}
\def\simlt{\mathrel{\spose{\lower 3pt\hbox{$\mathchar"218$}}
     \raise 2.0pt\hbox{$\mathchar"13C$}}}
\def\simgt{\mathrel{\spose{\lower 3pt\hbox{$\mathchar"218$}}
     \raise 2.0pt\hbox{$\mathchar"13E$}}}

\slugcomment{\it To be submitted to the Astrophysical Journal}

\shorttitle{Optical and mid-IR flares in SDSS1115+0544 }
\shortauthors{Yan et al.}

\begin{document}

\title{Rapid ``Turn-on'' of type 1 AGN in a quiescent early type galaxy SDSS1115+0544} 
\author{Lin Yan$^1$, Tinggui Wang$^2$, Jiang Ning$^2$, Daniel Stern$^3$, Liming Dou$^4$, C. Fremling$^5$, M.J. Graham$^5$, A. J. Drake$^5$, Chenwei Yang$^6$, K. Burdge$^5$, M. M. Kasliwal$^5$}
\affil{Caltech Optical Observatories, California Institute of Technology, Pasadena, CA 91125, USA, lyan@caltech.edu, 0000-0003-1710-9339 }
\affil{CAS Key Laboratory for Research in Galaxies and Cosmology, University of Science and Technology of China, Hefei, Anhui 230026, China; School of Astronomy and Space Science, University of Science and Technology of China, Hefei, Anhui 230026, China}
\affil{Jet Propulsion Laboratory, California Institute of Technology, 4800 Oak Grove Drive, Mail Stop 169-221, Pasadena, CA 91109, USA}
\affil{Center for Astrophysics, Guangzhou University, Guangzhou, 510006,
China}
\affil{Astronomy Department,California Institute of Technology, 1200 E. California Blvd, CA 91225, USA}
\affil{Polar Research Institute of China, 451 Jinqiao Road, Shanghai,
200136, China}

\begin{abstract} 
We present a detailed study of a transient in the center of SDSS1115+0544 based on the extensive UV, optical, mid-IR light curves (LC) and spectra over 1200\,days. The host galaxy is a quiescent early type galaxy at $z$\,=\,0.0899 with a blackhole mass of $2\times10^7M_\odot$.  The transient underwent a 2.5\,magnitude brightening over $\sim120$\,days, reaching a peak $V$-band luminosity (extinction corrected)  
of $-20.9$\,magnitude, then fading 0.5\,magnitude over 200\,days, settling into a plateau of $>$\,$600$\,days. Following the optical brightening are the significant mid-IR flares at $3.4$ and $4.5\mu$m, with a peak time delay of $\sim180$\,days. The mid-IR LCs are explained as the echo of UV photons by a dust medium with a radius of $5\times10^{17}$\,cm, consistent with $\rm E(B-V)$ of 0.58 inferred from the spectra. This event is very energetic with an extinction corrected $L_{bol}$\,$\sim$\,$ 4\times10^{44}$\,erg\,s$^{-1}$. Optical spectra over 400\,days in the plateau phase revealed newly formed broad H$\alpha, \beta$ emission with a FWHM of $\sim3750$\,km\,s$^{-1}$ and narrow coronal lines such as [Fe\,VII], [Ne\,V]. This flare also has a steeply rising UV continuum, detected by multi-epoch {\it Swift} data at $+700$ to $+900$\,days post optical peak.  The broad Balmer lines and the UV continuum do not show significant temporal variations.  The slow evolving LCs over 1200\,days, the constant Balmer lines and UV continuum at late-times rule out TDE and SN IIn as the physical model for this event. We propose that this event is a ``turn-on'' AGN, transitioning from a quiescent state to a type 1 AGN with a sub-Eddington accretion rate of $0.017M_\odot$/yr. This change occurred on a very short time scale of $\sim 120- 200$\,days. The discovery of such a rapid ``turn-on'' AGN poses challenges to accretion disk theories and may indicate such event is not extremely rare.

\end{abstract}

\section{Introduction}

In the past decade, time domain astronomy has made big strides forward with the advent of several wide area transient surveys, 
including the Panoramic Survey Telescope \&\ Rapid Response System \citep[Pan-STARRS;][]{Kaiser2002}, the Palomar Transient Factory \citep[PTF;][]{Law2009, Rau2009}, and the Catalina Real-Time Transient Survey \citep[CRTS;][]{Drake2009},  All-Sky Automated Survey for Supernovae (ASASSN) \footnote{http://www.astronomy.ohio-state.edu/$\sim$assassin/index.shtml},  The Asteroid Terrestrial-impact Last Alert System \citep[ATLAS;][]{Tonry2018} , and the most recently, Zwicky Transient Facility (ZTF; Graham et al. in prep, Bellm et al. in prep).  One particular area of the recent advances is transients detected in the centers of galaxies, especially quiescent galaxies where no previous nuclear activities were detected.  This type of transients includes both Tidal Disruption Events (TDEs) and Changing-Look AGNs (CLAGNs). The first phenomenon is the optical/UV flares generated by the disruption of a star when it gets too close to the central blackhole and was theoretically predicted in 1980s \citep{Hills1976,Rees1988,Phinney1989}.  The second type, CLAGNs, is due to the change of AGN accretion rates.  The changing states include  transformation from quiescent, non-accreting blackholes to active accreting type 1 or 2 AGNs \citep[e.g.][]{Gezari2017}, or spectral change between a type 2 and a type 1 AGN \citep[e.g][]{Ruan2016, Runnoe2016}, or periodic photometric variations (several years time scale) in some AGNs \citep[][]{Oknyanskij1978, Oknyanskij2007, Graham2015, Jun2015, Bon2016}.  Both TDEs and CLAGNs have become the focus of many recent studies \citep{Gezari2012, Yang2013, MacLeod2016, Ruan2016, Runnoe2016, Gezari2017, Yang2017, Arcavi2014, Holoien2014, Holoien2016, Blag2017,Oknyanskij2019}.  As the referee of this paper pointed out, spectral variations on time scales of years have been observed amongst nearby AGNs ({\it e.g} NGC4151, NGC3516) since 1960s \citep{Andrillat1968, Tohline1976,  Lyutyj1984, Penston1984}.  The renewed interests may lead us to deeper understanding this subject since we are now supplemented with more statistics and better data from large transient surveys.

Photometric surveys cannot immediately distinguish these two types of transients without additional follow-up spectroscopy.  This explains the recent surge of studies of CLAGNs, some are by-products from searches for TDEs.
Observational separations of TDEs and CLAGNs are not always clear \citep[e.g.][]{Blanchard2017} due to the similar fundamental physics governing both types of sources. This confusion is also due to the small numbers of TDEs and CLAGNs discovered so far. The increasing sample sizes in the future should help us build more complete empirical characterizations, thus also a coherent picture of accretion physics explaining both types of flaring events.

After a half dozen of optical/UV TDEs were discovered, it became apparent that the host galaxies of TDEs  are one specific type -- post-starburst galaxies with characteristic spectral features of E + A galaxies \citep{Arcavi2014, French2016}.  A hotly debated question is whether this is intrinsic or due to the TDE selection bias in the optical/UV wavelength.  One way to address this question is to examine if using infrared variability could be an effective means to identify TDE candidates because it is less biased against dusty star forming galaxies.  This motivated us to carry out a systematic search for mid-infrared flare sources using the data taken by the Wide-field Infrared Space Explorer \citep{Wright2010} from 2009 to 2017.  The detailed description of this search is prepared in a separate publication (Jiang et al. in prep). Here we summarize briefly the sample.  First, we impose four limits on our search -- (1) we focus only on the SDSS galaxies which have spectroscopic data. This gives us the critical information on the pre-flare properties of the host galaxies.  (2) we limit our search to $z<0.2$. This condition is very conservative, and can be easily extended in future studies.  (3) we only select sources whose mid-IR flares are still recent, {\it i.e.} their light curves (LC) are either still rising or have just turned over the peak, but not yet reached completely quiescent levels.  This criterion is chosen to give us sufficient time to take follow-up spectra.  (4) we focus only on variations on time scales $>$ 2 weeks. Therefore, we can coadd the single frame photometries taken within one week time span, thus reducing the photometric errors.  
All single frame source photometry is made available by the WISE and NEOWISE-R \citep{Mainzer2014} surveys through the IRSA data archive\footnote{https://irsa.ipac.caltech.edu/cgi-bin/Gator/nph-scan$?$mission=irsa\&submit=Select\&projshort=WISE}.  We found a total of 28 galaxies with mid-IR flares with a magnitude change $>$ 0.5\,magnitude (Vega).  As shown in the histogram of WISE magnitude variation in Figure 1 of \citet{Stern2018}, such a brightness change is significant, indicative of intrinsic physical variation.  The sample light curves and the selection details are presented in Jiang et al. (in prep).

This paper is dedicated to a single object SDSS\,J111536.57+054449.7 (hereafter SDSS1115+0544), which has a rich suite of data, including optical and mid-IR light curves, a spectral sequence over a time scale $>1$year, and new {\it Swift} observations.   
As our analyses show below,  this object may represent a class of early type galaxies, rapidly transforming from quiescent states over only a few hundred days to actively accreting type 1 AGNs.  We demonstrate that there are some ambiguities between this type of transients and peculiar type II supernovae, and larger samples of such events over several years would help to firmly establish the classification criteria. 
\begin{figure}[!ht]
\center
\includegraphics[width=0.98\linewidth]{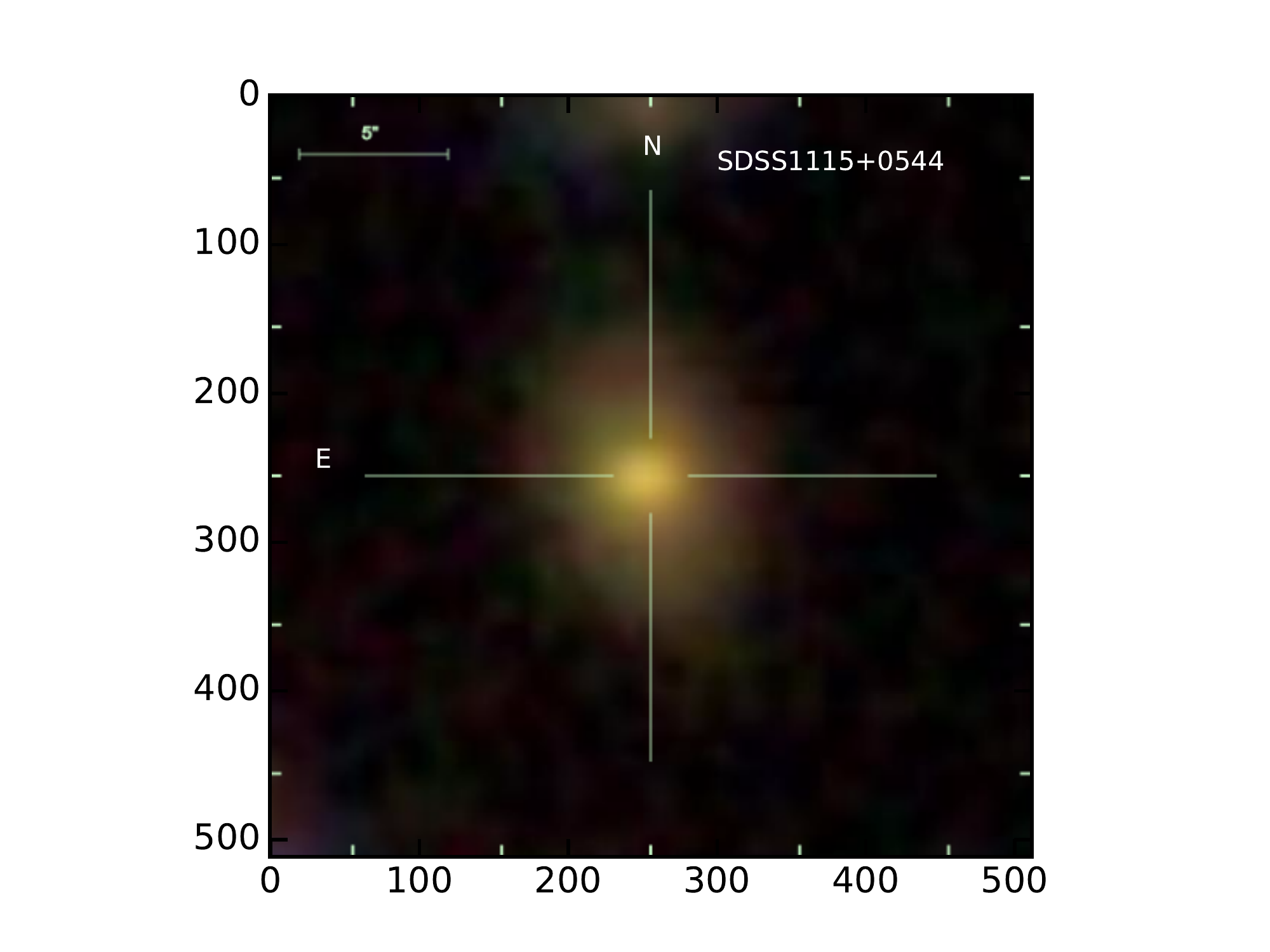}
\includegraphics[width=0.98\linewidth]{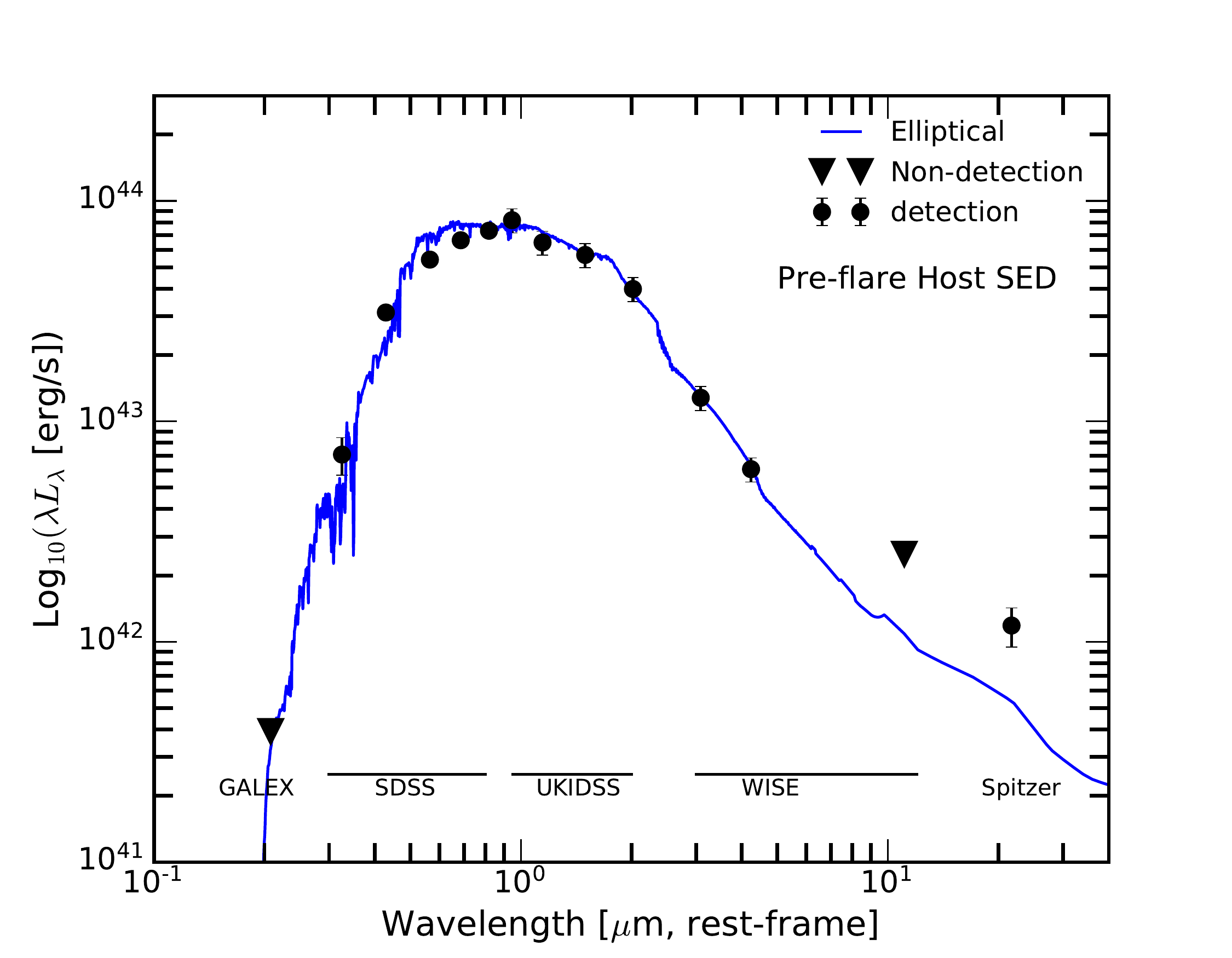}
\caption{ The two-panel plot shows the composite image, illustrating the morphological properties, and the spectral energy distribution of the host galaxy SDSS1115+0544 in its quiescent state.  The {\it Spitzer} 24$\mu$m flux is in excess of the quiescent early type galaxy SED template, indicating dust obscured star formation, which is not accounted for by the model SED template. \label{fig:host}}
\end{figure}

\section{The quiescent state of the host galaxy }
\label{sec:basicinfo}
SDSS115+0544 has the sky position of RA\,=\,11:15:36.57 and DEC\,=\,05:44:49.73 Equinox J2000.0 and at redshift of 0.08995.  Before SDSS1115+0544 displayed optical and infrared flare in 2015,  its quiescent state has multi-band photometry in $ugriz$ (SDSS), $YJHK$ (UKIDSS), and 3.4 ($\rm W1$), 4.6 ($\rm W2$), 12 ($\rm W3$, not detected) and 12$\mu$m ($\rm W4$, not detected) (ALLWISE) and 24$\mu$m ({\it Spitzer}). 
SDSS1115+0544 is a fairly bright, bulge dominated galaxy with $g$ \&\ $r$ magnitude of 17.96 and 17.06\,mag (AB) respectively. Its pre-flare, panchromatic photometry is shown in the bottom panel of Figure~\ref{fig:host},  and at the optical/near-IR wavelength, is well fit by the spectral energy distribution (SED) template of an elliptical galaxy \citep{Polletta2007}.  In its quiescent state, the host galaxy was not detected in the near-UV (2267\AA) by GALEX ($>23.5$\,mag), in the WISE W3 and W4 bands, and in the FIRST 20cm survey ($1\sigma$\,=\,0.15\,mJy).

\begin{figure*}[!t]
\center
\includegraphics[width=0.99\linewidth]{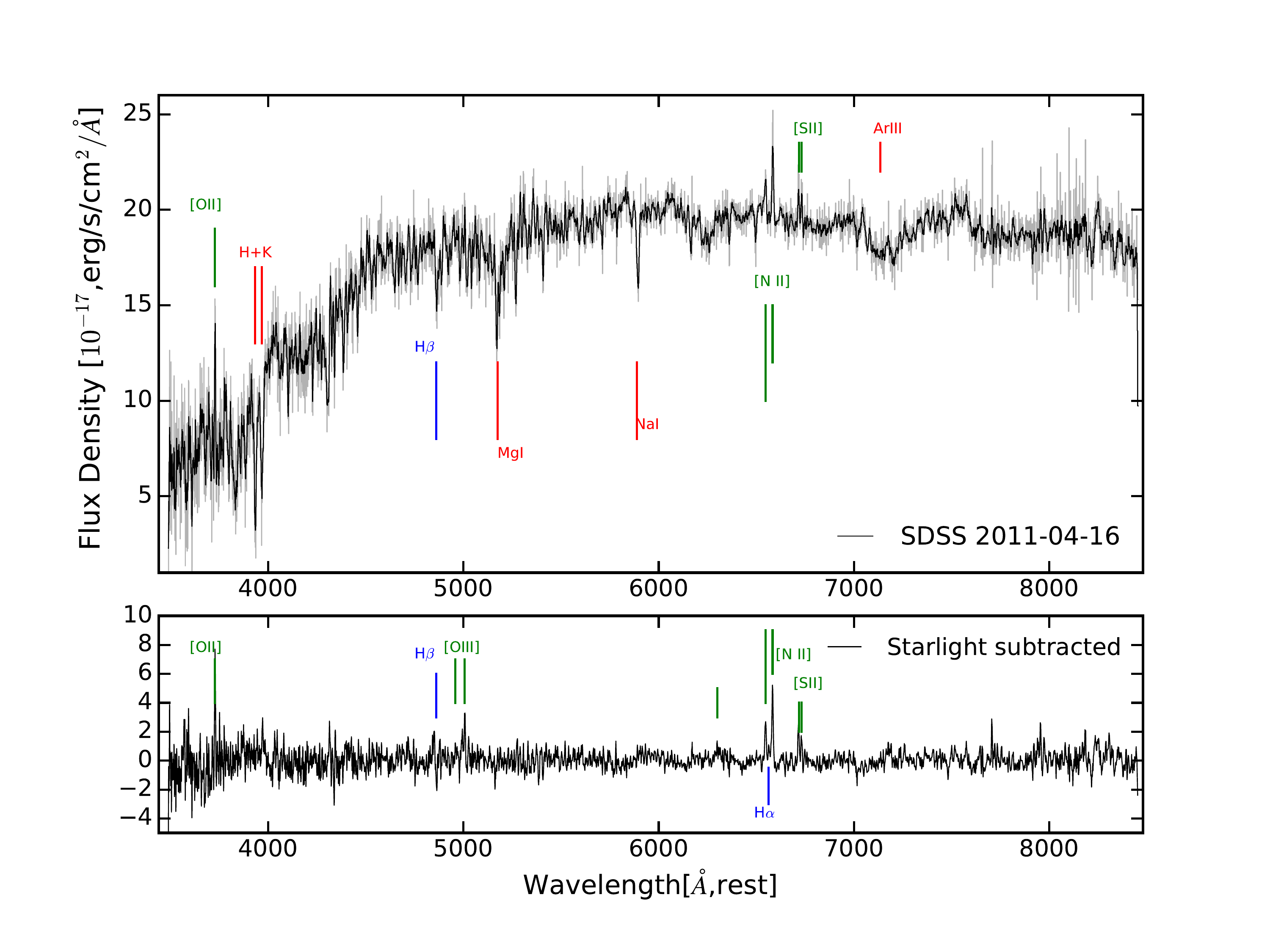}
\caption{The two-panel figure presents the optical spectrum in the quiescent state taken by SDSS (top) and the star light subtracted spectrum (bottom), illustrating the absorption lines and weak emission lines from the host galaxy.  The line identifications are marked with vertical bars. \label{fig:host2}}
\end{figure*}

The pre-flare SDSS spectrum is shown in Figure~\ref{fig:host2}, characteristic of a passive evolving early type galaxy, with strong absorption lines such as Ca\,H+K, Ca\,I, Mg\,I\,5175\AA\ from a dominant old stellar population. The spectral models from the SDSS DR14 show that this galaxy has no current star formation activity. However, SDSS1115+0544 is detected by {\it Spitzer} at 24$\mu$m in 2006 and 2007.  The dust obscured star formation rate (SFR) could be as much as $\rm 0.2M_\odot$/yr based on the calibration relation $\rm SFR (M_\odot/yr) = 2.04\times10^{-43}\times \nu L_{\nu}(24\mu m, erg/s)$ \citep{Calzetti2013}.  In comparison with the Milky Way $\rm SFR \sim 0.68 - 1.45M_\odot$/yr \citep{Robitaille2010},  SDSS1115+0544 has a lower star formation rate.  

The SDSS DR14 archive lists several derived parameters based on absorption lines, specifically, the velocity dispersion $\sigma_v$ is $128 \pm17$\,km\,s$^{-1}$ (Wisconsin method, Maraston models) and the stellar mass $M_{star}$\,=\,$3.47^{+0.42}_{-0.52}\times10^{10}M_\odot$ based on the Portsmouth method (passive model).  Assuming the M$-\sigma_v$ relation of $\rm log_{10} (M_{BH}, M_\odot) = 8.32 + 5.64 log_{10}(\sigma / 200)$ \citep{McConnell2013}, we infer the blackhole mass of $\rm log_{10}(M_{BH}) =  7.29^{+0.18}_{-0.11}$.

Besides absorption lines, SDSS115+0544 also displays weak nebular emission lines. To measure the reliable emission line ratios, we subtract the synthetic star light spectrum, made by the FIREFLY models for the eBOSS spectra (SDSS DR14) (Figure~\ref{fig:host2}) \citep{Comparat2017} \footnote{https://www.sdss.org/dr14/spectro/eboss-firefly-value-added-catalog/}.  The star light subtracted spectrum reveals [O\,III]\,4959,5007\AA\ and [N\,II]\,6548,6583\AA\ doublets, but very little H$\beta$ and H$\alpha$, yielding line ratios of $\rm Log_{10}(\frac{[O\,III]}{H_\beta}) > 0.25$ and $\rm Log_{10}(\frac{N\,II}{H_\alpha}) = 0.23$.  For comparison, star forming galaxies generally have $\rm Log_{10}(\frac{N\,II}{H_\alpha}) < -0.2$, and strong AGNs have $\rm Log_{10}(\frac{[O\,III]}{H_\beta}) > 0.5$.  SDSS1115+0544 falls within the lower-right region of the $\rm [O\,III]/H_\beta - N\,II/H_\alpha$ diagram, {\it i.e.} Baldwin, Phillips \&\ Telervich (BPT) diagram. \citet{Kauffmann2005} classification of LINERs have $\rm Log_{10}(\frac{[OIII]}{H\beta})< 0.477$ and $\rm Log_{10}(\frac{[NII]}{H\alpha}) > -0.22$.
The very weak H$\beta$ and H$\alpha$ lines and the corresponding line ratios indicate that SDSS1115+0544 is aligned with the definition of a Low Ionization Nuclear Emission Line Region (LINER), based on the diagnostics by \citet{Kewley2006,Cid2010}.  
The SDSS DR14 archive also classified this object as a LINER.

The physical nature of the emission lines from a LINER is still a subject of debate.  Some LINERs are thought to be weak AGNs. However, some emission lines in some LINERs are thought to be mainly produced by evolved stars during a short but very energetic post-AGB phase, as shown by a study of integral field spectroscopy of nearby LINERs \citep{Singh2013}.  We will discuss the implications of the transient flare in a LINER in \S~\ref{sec:discuss}.

\section{Transient Observations in SDSS1115+0544 }
\label{sec:data}

\subsection{Light Curves}
SDSS1115+0544 has optical light curve from CRTS \citep{Drake2009}, between January 2006 to May 2017.  The photometry comes from both the 1.5\,meter Mt. Lemmon Survey telescope and the 0.7\,meter Catalina Sky Survey telescope.  We perform image subtraction analyses using the custom software, Fpipe \citep{Fremling2016}. The reference images are constructed using the CRTS images taken during the quiescent phase (before 2015). The transient photometry is measured by PSF fitting and calibrated on to the standard Johnson-Cousins $V$-band using SDSS photometry of the stars within the same images.  Our optical photometry is in AB magnitude.

\begin{figure}[!t]
\center
\includegraphics[width=3.7in,height=5in]{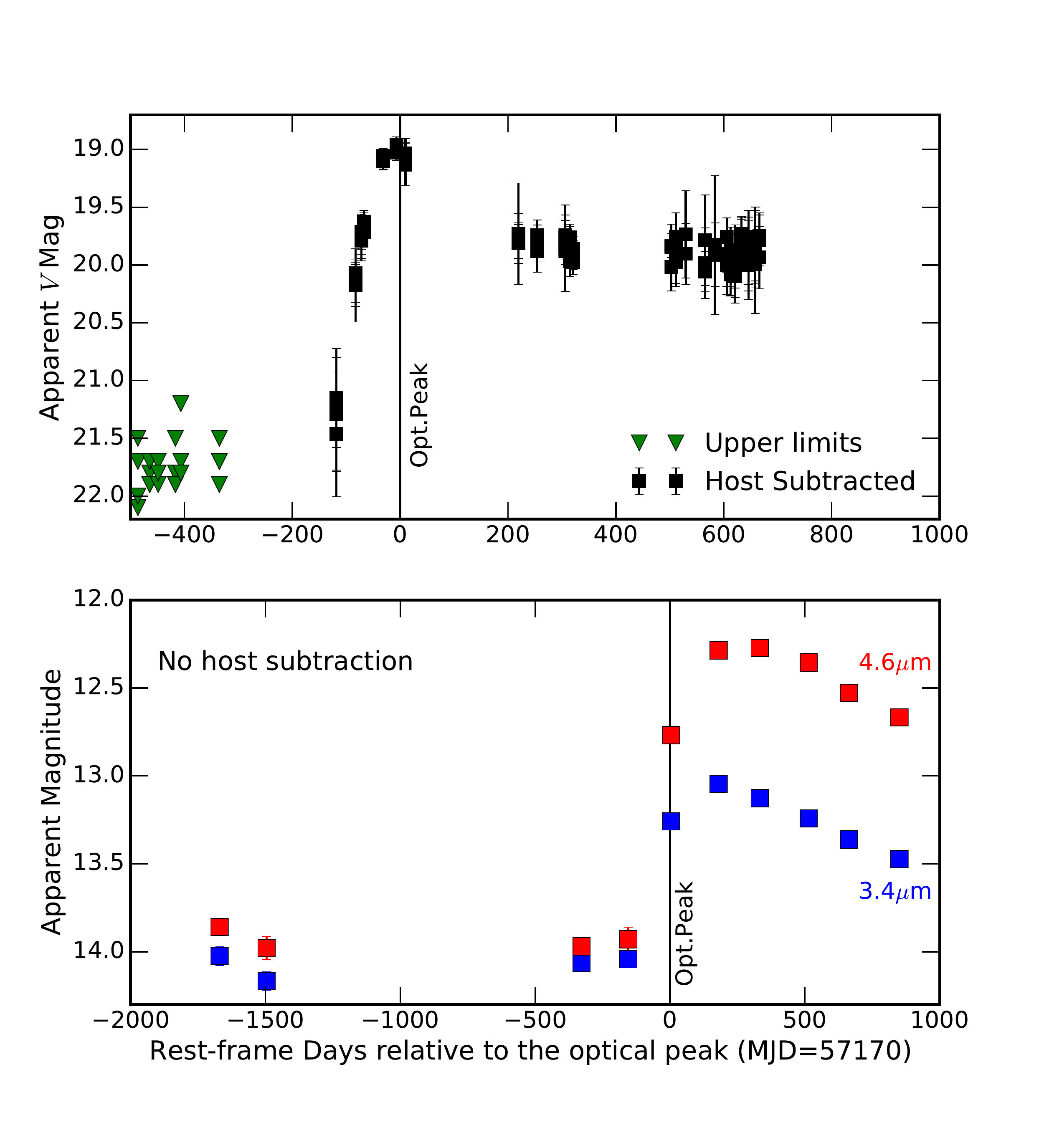}
\caption{The observed (apparent magnitudes) light curves (LC) in the optical $V$ band (top) and the mid-infrared $3.4\mu$m and $4.6\mu$m (bottom).  The optical LC is host subtracted (see details in the text) and the MIR LCs are the total apparent magnitudes. \label{fig:rawlc}}
\end{figure}

The mid-infrared photometry is from WISE \citep{Wright2010} and NEOWISE \citep{Mainzer2014} catalogs, covering a time span of 2010 to 2018.  There's a gap between 2011 and 2013 because the WISE space telescope was placed in hibernation after the exhaustion of its cryogen at the end of the primary mission. Except this gap, the WISE has visited SDSSJ1115+0544 at W1 ($3.4\mu$m) and W2 ($4.6\mu$m) every half year, with $10-20$ single exposures from each visit.  We simply retrieved the W1 and W2 PSF profile-fit magnitudes from the catalogs, assuming that SDSS1115+0544 is a point-like source in the WISE images. We have searched the catalogs and there are no photometric contaminations from other sources within $10^{"}$ of our target. Data with poor quality frame ($qi_{fact}\ge0.5$), charged particle hits ($saa_{sep}<0$), scattered moon light ($moon_{masked}=1$) and artifacts ($cc_{flags}>0$) are rejected. The MIR photometry for SDSSJ1115+0544 is distributed over 10 epochs with intervals of $\sim$ six months. No reliable short time scale variability is detected within each individual epoch and thus we have binned the data every half year and computed the median values following our previous work \citep{Jiang2016, Jiang2017}. The magnitudes in the first four epochs show little change. However, the fifth epoch presents a sudden brightening of $>$0.5 magnitudes in both bands (Figure~\ref{fig:rawlc}). The light curves reach their peak at the sixth epoch and then decline. 
The mean magnitudes of the first four epochs are $14.083\pm0.028$ and $13.91\pm0.041$\,magnitude at $3.4\mu$m and $4.6\mu$m respectively. The $\rm [W1 - W2]$ color in the quiescent state is 0.173, $<$0.8\,magnitude, above which is the selection for mid-IR AGNs \citep{Stern2012, Yan2013}. This suggests that SDSS1115+0544 is a normal, non-AGN host in its quiescent state.  We note here that mid-IR magnitudes are all in Vega system as provided by the WISE archive. 

Figure~\ref{fig:rawlc} displays the optical (top) and mid-IR LCs (bottom) covering a time range of $\sim$\,12000\,days.  Here the apparent magnitudes are used to illustrate the brightness changes in the observed frame. The photometric flare in SDSS1115+0544 is significant and well captured by the CRTS and WISE observations.

Additional photometric data is from the {\it Swift}, including both UV Optical
Telescope (UVOT) and X-ray Telescope (XRT).  SDSS1115+0544 is not detected in the UV by GALEX before the flare, and after the 2015 flare, it is significantly detected by {\it Swift}. 
The three epochs of {\it Swift} UV photometry were obtained on UT 2017-07-02, 2017-07-13 and 2017-12-26, see Table~\ref{tab:swiftphot} for the details, which are at the tail end of optical and mid-IR flares.  The data were reduced using the standard software tool, {\it uvotimsum}, provided by High Energy Astrophysics Software (HEASoft) from NASA Goddard Space Flight Center (NGFS). The photometry is measured using a $3^{\prime\prime}$ radius aperture, which is corrected out to $5^{\prime\prime}$ in radius for the total flux.  The {\it Swift} XRT data was taken on the same three dates,  and no detections were found at the position of SDSS1115+0544.  Coadding all of the XRT data with a total of 4.27\,kseconds, we derived a $1\sigma$ upper limit of $7\times10^{-14}$\,erg\,s$^{-1}$\,cm$^{-2}$ in the {\it Swift} X-ray band.

\begin{figure*}[!t]
\center
\includegraphics[width=1.1\linewidth]{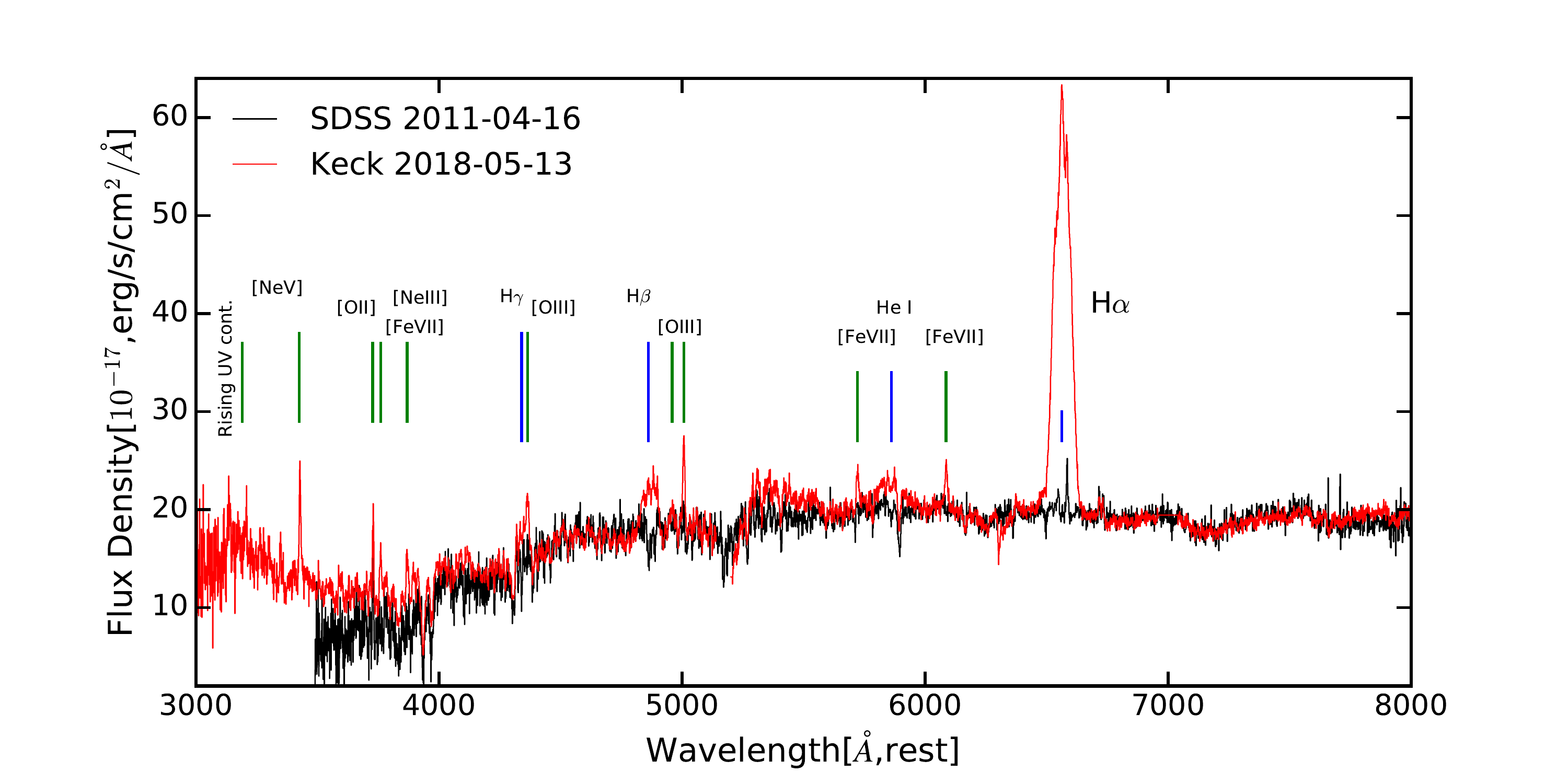}
\caption{The observed Keck spectrum on UT 2018-05-13 (red) in comparison with the quiescent state SDSS spectrum from UT 2011-04-16 (black).  The prominent rising UV continuum and new emission lines are marked. We note that the Keck spectrum shown here is made available for download as ascii tables in this paper. \label{fig:rawspec}}
\end{figure*}

\subsection{Spectroscopy}
The earliest spectrum is from Sloan Digital Sky Survey (SDSS) taken in UT 2011-04-16. After discovering the strong mid-infrared flare in early 2017, we began our spectroscopic monitoring through UT 2018-05-13.  The optical spectra were taken with both the Double Spectrograph \citep[DBSP,][]{Oke1983} mounted on the Hale 200\,inch telescope at Palomar Observatory as well as the Low Resolution Imager and Spectrograph \citep[LRIS,][]{Oke1995} on the Keck 10\,meter telescope.  These observations were taken under various different conditions, thus extra cares are taken when we compare line fluxes from different epochs as in \S~\ref{sec:spchange}.

The DBSP observations used the D550 dichroic which splits the incoming light at 5500\AA\ into the blue and red beam for two different grism/grating.  The pixel scales are $0.293^{\prime\prime}$ and $0.398^{\prime\prime}$ per pixel for the red and blue CCD respectively.  The slit width of either $1^{\prime\prime}$ or $1.5^{\prime\prime}$ was used depending on the seeing condition.  All of the DBSP spectra were reduced using a python script written by Eric Bellm (University of Washington) and Kevin Burdge (Caltech).  The spectra are flux calibrated using standard star observations from the same night.  
The Keck LRIS spectroscopy used the D560 dichroic in combination with the blue grism of 600\,line per mm \&\ blazed at 4000\,\AA\  and the red grating of 400\,line per mm blazed at 8500\,\AA. The long slit has a width of 1$^{\prime\prime}$. This setup achieved a spectral resolution of $6.5 - 7.1$\,\AA\ for the full wavelength range of $3300 - 9500$\,\AA.
The observation was taken under excellent weather, with seeing of $0.6^{\prime\prime}$ and clear photometric conditions.  The LRIS data was reduced by co-author Daniel Stern using his well tested pipeline from previous studies.  
Table~\ref{tab:speclog} shows the technical details of these observations. 

Figure~\ref{fig:rawspec} compares the pre-flare (UT 2011-04-16 SDSS) and post-flare (UT 2018-05-13 LRIS) optical spectra, revealing striking newly formed features, including the rising UV continuum, prominent broad H$\alpha$ emission lines as well as high ionization metal emission lines. The post-flare Keck spectrum is available for download as ascii tables as the part of this paper. See full discussions in \S~\ref{sec:spchange}.

\section{Analyses and Results \label{sec:results}}
\begin{figure}[!ht]
\center
\includegraphics[width=3.7in,height=7in]{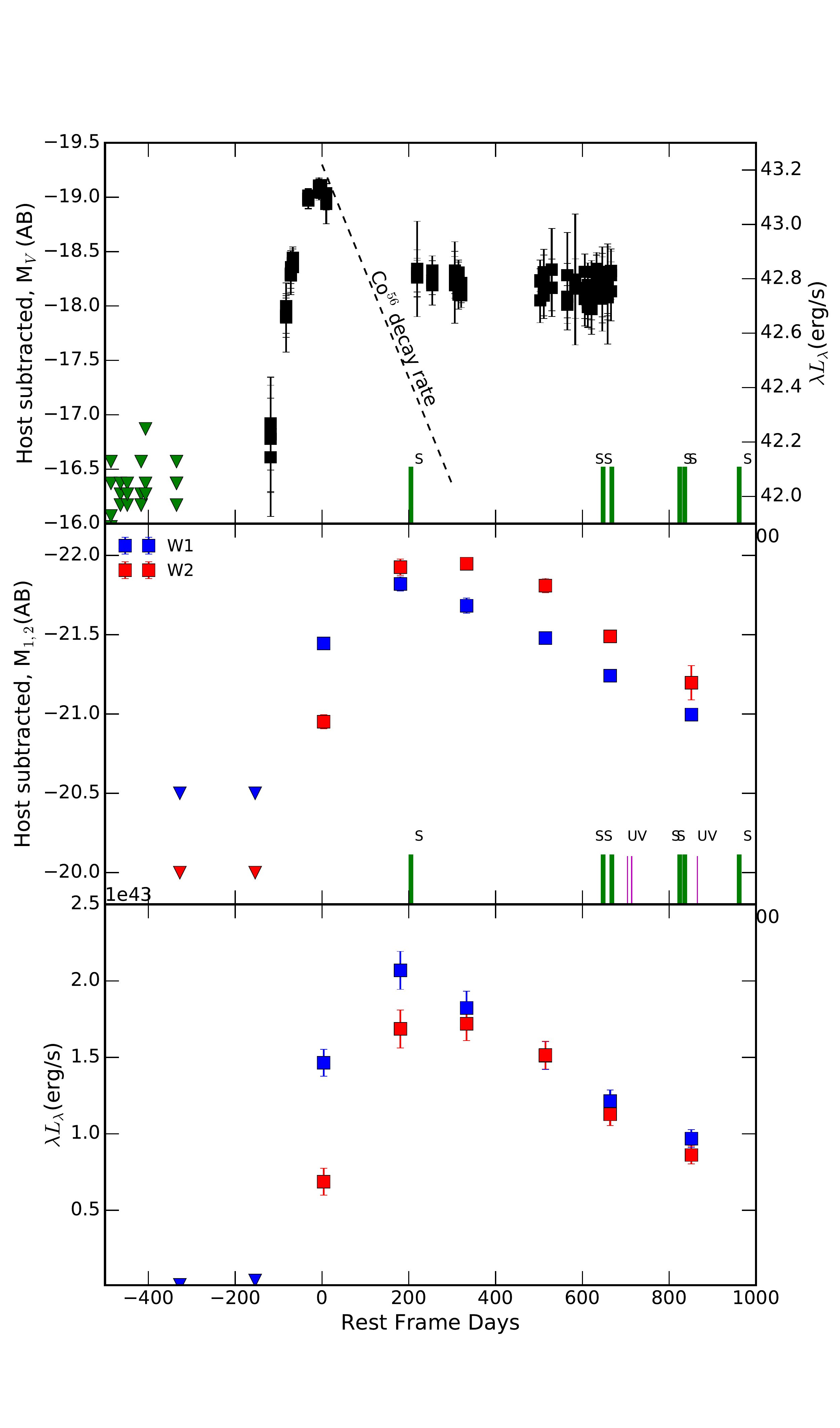}
\caption{Optical and mid-infrared light curves of SDSS1115+0544 presented in host subtracted absolute magnitudes (AB) and luminosities over a time span of $\sim1200$\,days. The downward triangle symbols are the non-detection limits. The vertical green bar with a letter "S" marks the epoch with the spectroscopic data.  In addition, the vertical purple bar marks the time when {\it Swift} UV photometry was taken. Dashed line marks the fading slope of radioactive decay rate from Co$^{56}$. \label{fig:lc}}
\end{figure}

We convert the temporal variations in apparent brightness (Figure~\ref{fig:rawlc}) to host-subtracted absolute magnitudes and monochromatic luminosity $\lambda L_{\lambda}$. The results are shown in Figure~\ref{fig:lc}. For both optical and mid-IR, the host-subtracted light curves are available for download in ascii tables. We release the apparent magnitudes as functions of time with the explanatory information included in the headers of the ascii tables. The absolute magnitudes are calculated from $\rm m - M = 5log_{10}(D_L/10pc) + KC$, where $\rm D_L$\,=\,414\,Mpc is the co-moving luminosity distance and $\rm KC$ is the K-correction. At $z=0.0895$, $\rm KC$ is relatively small ($\sim0.1$\,magitude) \citep[e.g.][]{Hogg2002}, and ignored by our analysis since our mid-IR flares are not measured from image subtractions, thus the systematic uncertainties in the MIR light curves could be as large as 0.05\,magnitude.
Mid-IR Vega magnitudes are converted to AB system to be consistent with that of optical LC. The transformation between $\rm [W1,W2](vega)- [W1,W2](AB)$\,=\,$[-2.65,-3.29]$ \footnote{http://wise2.ipac.caltech.edu/docs/release/allsky/expsup/}.  The conversion factor in $\rm W2$ is large, which explains why the flare has smaller $\rm W1-W2$ color in AB magnitude than in the Vega system.

In addition, the apparent magnitudes are converted into flux densities $f_{\lambda}$ in erg\,s$^{-1}$\,cm$^{-2}$\,\AA$^{-1}$ using the zero points tabulated in the WISE Explanatory Supplement.
The mid-IR photometry at the first four epochs does not vary and the galaxy is in the quiescent state (Figure~\ref{fig:rawlc}). The mean magnitudes of these four epochs are taken to represent the host galaxy emission at the level of $\rm <W1>$\,=\,14.075 and $\rm <W2>$\,=13.934\,mag respectively. These host emissions are removed to produce the host-subtracted light curves. 
The three panels in Figure~\ref{fig:lc} illustrate several important results, which are discussed in the following subsections.

\subsection{Mid-Infrared Time Lag}
\label{sec:timelag}

The mid-Infrared flare is clearly delayed by about $\sim180$\,days ($\Delta t_{delay}$) relative to the optical flare (peak-to-peak).
Independent of the physical processes of the central UV/optical flare, the mid-Infrared LCs  are likely the product of the dust echo of the UV/optical flares.  The basic idea is that UV/optical photons travel a distance $R$,  and get absorbed by the dust grains, which in turn re-radiate the excess energy in the infrared.  Roughly speaking, the time delay $\Delta t_{delay}$ is  $\sim R/c$, where $c$ is the speed of light.  $\Delta t_{delay} \sim 180$\,days implies that the dusty medium is at a distance of $\sim5\times10^{17}$\,cm from the UV/optical source.  For comparison, the dust sublimation radius is $R_{sub} (pc) = 
1.3(L_{UV}/10^{46}{\rm \,erg\,s^{-1}})^{0.5} (T_{sub}/1500\,K)^{-2.8}\,(a/0.05\mu m)^{-0.5}$,
with $a$ the size of dust grain and $T_{sub}$ the grain sublimation temperature \citep{Barvainis1987}. With the extinction corrected $L_{UV}\sim1\times10^{44}$\,erg\,s$^{-1}$ (see \S~\ref{sec:flaresed}), and assuming graphite dust grains (their $T_{sub} = 1500$\,K and $a=0.05\mu m$), we have the sublimation radius $R_{sub}$\,=\,0.13\,pc\,=\,$4\times10^{17}$\,cm, which is comparable to the radius of the dust shell computed from the time lag between the optical and mid-IR LCs.
 
 

The optical and mid-IR LCs also constrain the geometry, orientation and thickness of the dusty gas.  For example, the fact that optical and UV flare signals are observed can rule out  the spherical dust distribution with high filling factors (or high optical depths).
Detailed modelings of the LC shape can constrain the physical parameters of the dusty medium. One simple example is the tilted dust ring which is inferred from the mid-IR flare in IRAS F01004$-$2237  \citep{Dou2017}.  For our sample of mid-IR flares, the LC modeling is planned for a separate paper (Wang et al. in prep). 

\begin{figure*}[!t]
\includegraphics[width=1\linewidth]{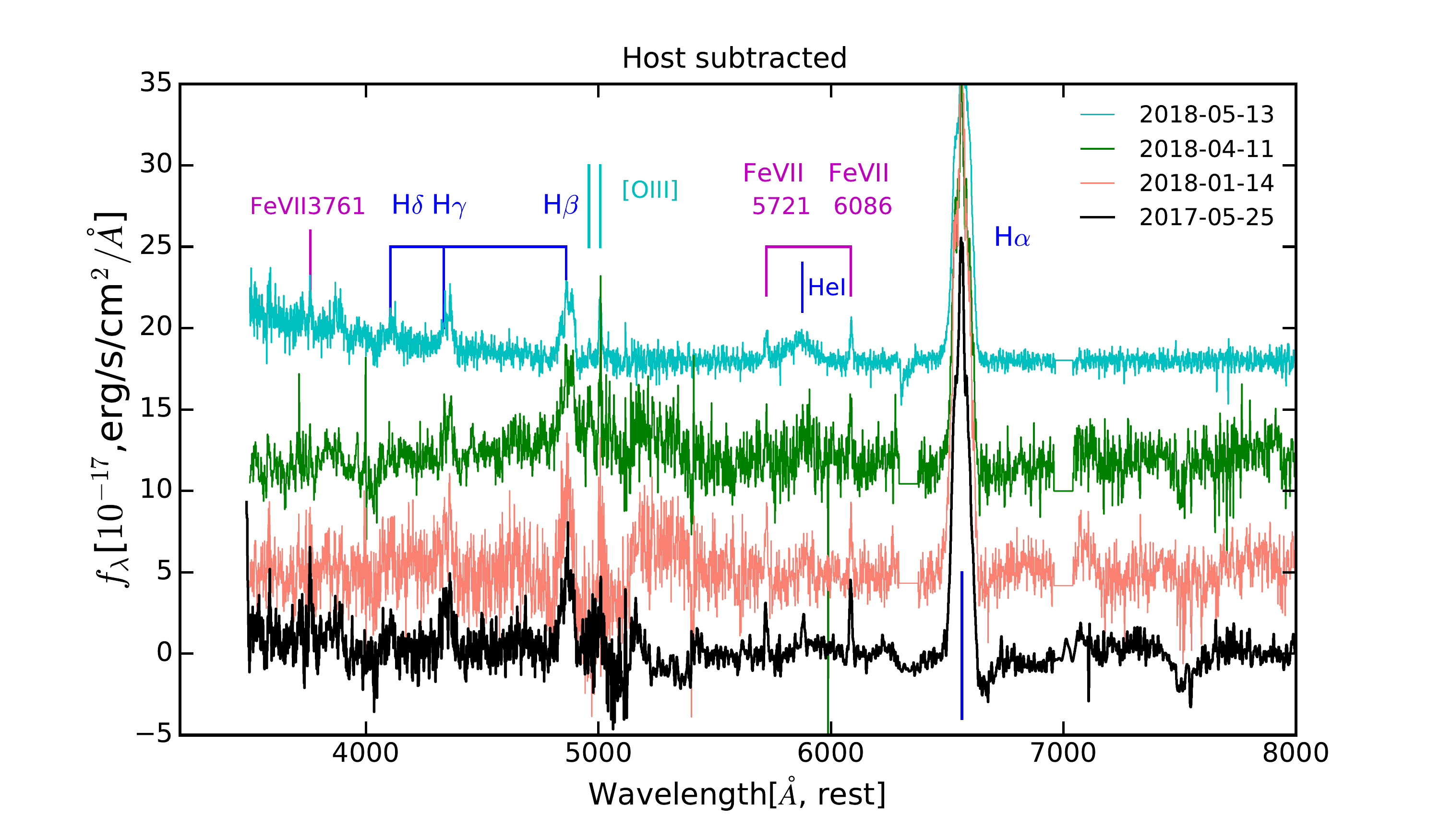}
\caption{ The host galaxy subtracted spectra are displayed. The data are from four epochs in $2017 - 2018$, after the peak of the optical flare. All prominent new spectral emission lines are marked with vertical bars. \label{fig:subspec}}
\end{figure*}

\subsection{Dramatic spectral changes}
\label{sec:spchange}

A rich set of new spectral features emerged in the post-flare optical spectra (Figure~\ref{fig:rawspec}). This includes strong \&\ broad H$\alpha$, H$\beta$ and He\,I\,5875\,\AA, steeply rising UV continuum ($\lambda>3500$\AA), narrow metal lines, and most interestingly [Fe\,VII] coronal lines.  Inspection of the 2-dimensional spectra found that H$\alpha$ is from the central spatially unresolved region. This rules out supernova exploded at a distance $>1.5^{\prime\prime}$ from the center (angular size distance of 2.5\,kpc). 

Six optical spectra were obtained between UT 2017-05-24 and 2018-05-13 (Table~\ref{tab:speclog}).   
The P200 spectra were extracted using an aperture of $4^{"}$, roughly $2.5 - 3$ times of the typical seeing. This is large enough to capture most of the line emission from the central point source.  For comparison, the Petrosian radius containing 50\%\ of the total flux is $1.8^{\prime\prime}$ for the host galaxy SDSS1115+0544. Each flux calibrated spectrum
is matched to the SDSS spectral continuum and Ca H+K absorption features to produce the host galaxy subtracted, flare spectrum.  Figure~\ref{fig:subspec} displays the host subtracted spectra from four different epochs. Broadly speaking, these flare spectra do not show prominent variation during one year period from $2017 - 2018$.  The last epochal spectrum was taken with LRIS on the Keck under extremely good condition, with clear sky and $0.6^{\prime\prime}$ seeing.  The Keck spectrum was extracted with a $1.5^{\prime\prime}$ ($2.5\times$\,seeing) aperture, capturing most of the H$\alpha$ emission from the central unresolved region, with the total line flux of $2.1\times10^{-14}$\,erg\,s$^{-1}$\,cm$^{-2}$, thus $L_{H\alpha}$\,=\,$4\times10^{41}$\,erg\,s$^{-1}$.

All emission line measurements are made with the host subtracted spectra. Table~\ref{tab:metaline} tabulates the line wavelengths, integrated fluxes and line widths (FWHM) measured from the Keck spectrum (UT 2018-05-13). 

\subsubsection{High dust content and the inferred $E(B-V)$}
In its quiescent state, SDSS1115+0544 was detected by {\it Spitzer} at 24$\mu$m, with the flux in access to the SED of a pure early type galaxy, as shown in Figure~\ref{fig:host}. This mid-infrared excess suggest that the galaxy contains a substantial amount of warm dust. 

The UV/optical flare ionized neutral hydrogen, producing Balmer recombination lines. 
In the Keck spectrum, the H$\alpha$-to-H$\beta$ line ratio is 5.45, significantly larger than 3.1, a value commonly found for blue AGNs \citep{Dong2008}.    
We adopt the relation derived by \citet{Caplan1986},   $A_{H\alpha}$\, =\, $5.25 [ {\rm log_{10}} (H\alpha/H\beta) - {\rm log_{10}}(2.86t_e^{-0.07}) ]$, where H$\alpha$/H$\beta$ is the line flux ratio, $t_e$ is the electron temperature in units of $10^4$\,K.  The standard extinction curve with $R_V = A_V/E(B - V) = 3.1$ \citep[Table~7.1][]{Osterbrock2006} gives $A_{H\alpha} / A_V = 0.82$. Combining these two equations, we derive $E(B - V)$\,=\,0.58 and $A_V$\,=\,1.8\,magnitude. The extinction corrected $V$-band peak luminosity is $\nu L_\nu (V) = 7\times10^{43}$\,erg\,s$^{-1}$, making it comparable to that of most luminous supernovae and TDEs.

Dust absorption can also explain the {\it Swift} X-ray non-detections. 
The flux limit is $7\times 10^{-14}$\,erg\,s$^{-1}$\,cm$^{-2}$ at the 90\%\ confidence,  corresponding to the luminosity limit of $1.4\times10^{42}$\,erg\,s$^{-1}$.  A previous study of a sample of 63 AGNs 
\citep{Gelbord2009} has found that the soft X-ray luminosity is correlated with the coronal line luminosity, with $\rm Log_{10}[{\it  f_{\rm xray}/f}([Fe\,VII])]$\,=\,$(3.37\pm0.51)$. This sets the expected intrinsic soft X-ray luminosity to $(0.5 - 5.3)\times10^{43}$\,erg\,s$^{-1}$, which is higher than the {\it Swift} non-detection limit by a factor of $(3 - 30)$.  Non-detection in X-ray can be explained if the expected intrinsic soft X-ray flux is absorbed by a gas column density of $3\times10^{21}$\,cm$^{-2}$.

\begin{figure}[!ht]
\includegraphics[width=1\linewidth]{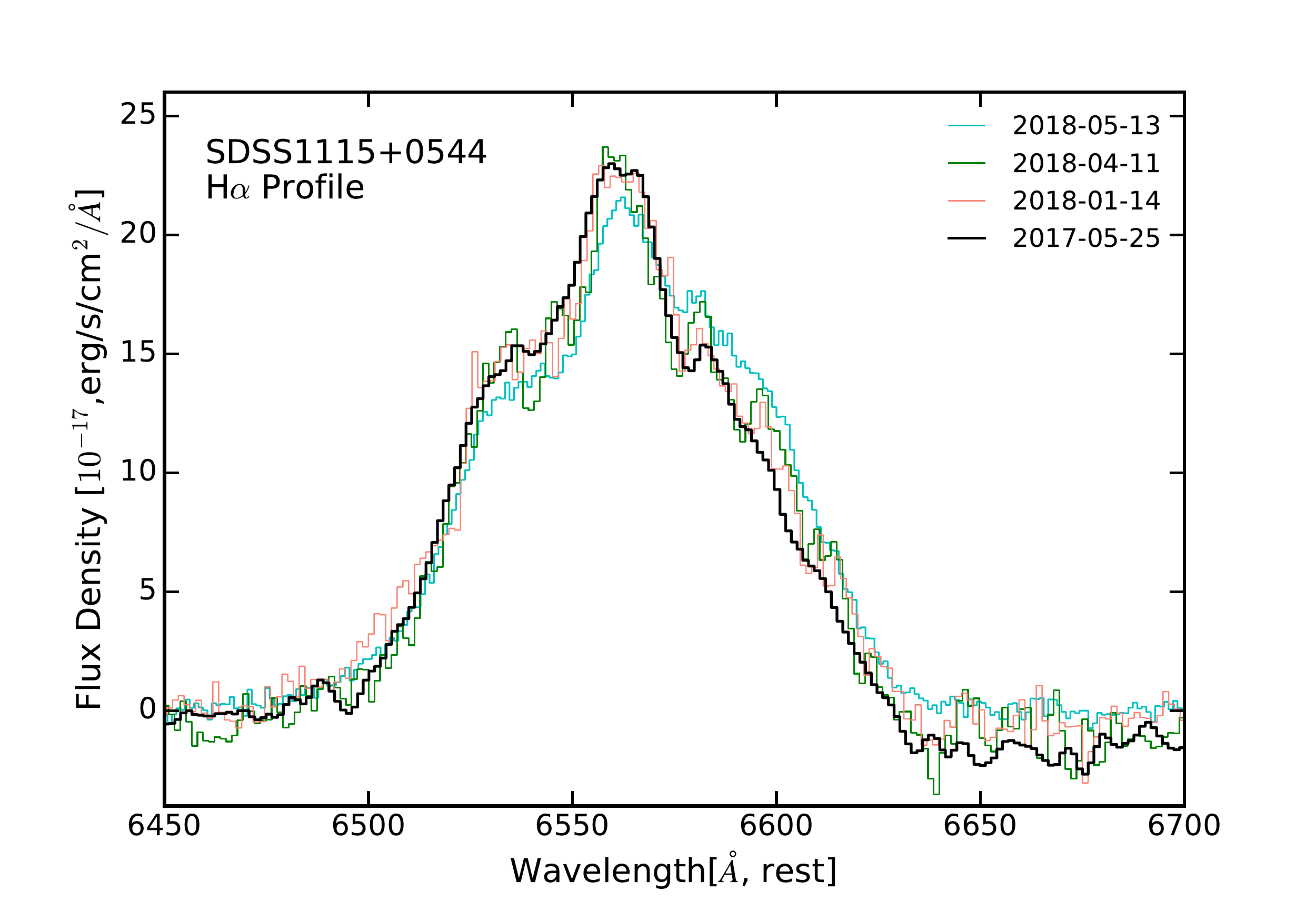}
\caption{The zoom-in display of the H$\alpha$ spectral region from the P200 spectra at four epochs spanning one year. \label{fig:ha}}
\end{figure}

\subsubsection{Temporal behavior -- constant broad H$\alpha$ line and variable narrow [O\,III]\,5007 \&\ [Fe\,VII]\,6086\AA\ lines} 
We measured the broad H$\alpha$ line fluxes using the four epochal P200 spectra, and do not find significant temporal variations, with the line fluxes  
in a narrow range, $(1.9, 2.7, 1.8, 2.0)\times10^{-14}$\,erg\,s$^{-1}$ for UT 2018-04-11, 2018-01-14, 2017-11-24, and 2017-05-24 respectively.  These values are similar to the integrated flux measured from the Keck spectrum as well. In addition, SDSS1115+0544 was observed by LAMOST (spectral fiber diameter $3^{\prime\prime}$) on UT 2016-01-06 \citep{Yang2017}. With the same method, we measured the H$\alpha$ line flux of $2.0\times10^{-14}$\,erg\,s$^{-1}$.  We infer $L_{H\alpha}\sim4\times10^{41}$\,erg\,s$^{-1}$.  The broad H$\alpha$ FWHM has a velocity width of $\sim 3750$\,km\,s$^{-1}$ after corrected for the instrumental resolution. Similar to the constant line flux, the line width does not change with time.  He\,I\,5876\AA\ and H$\beta$ line widths are consistent with that of H$\alpha$.   

\begin{figure}[!ht]
\includegraphics[width=1.15\linewidth]{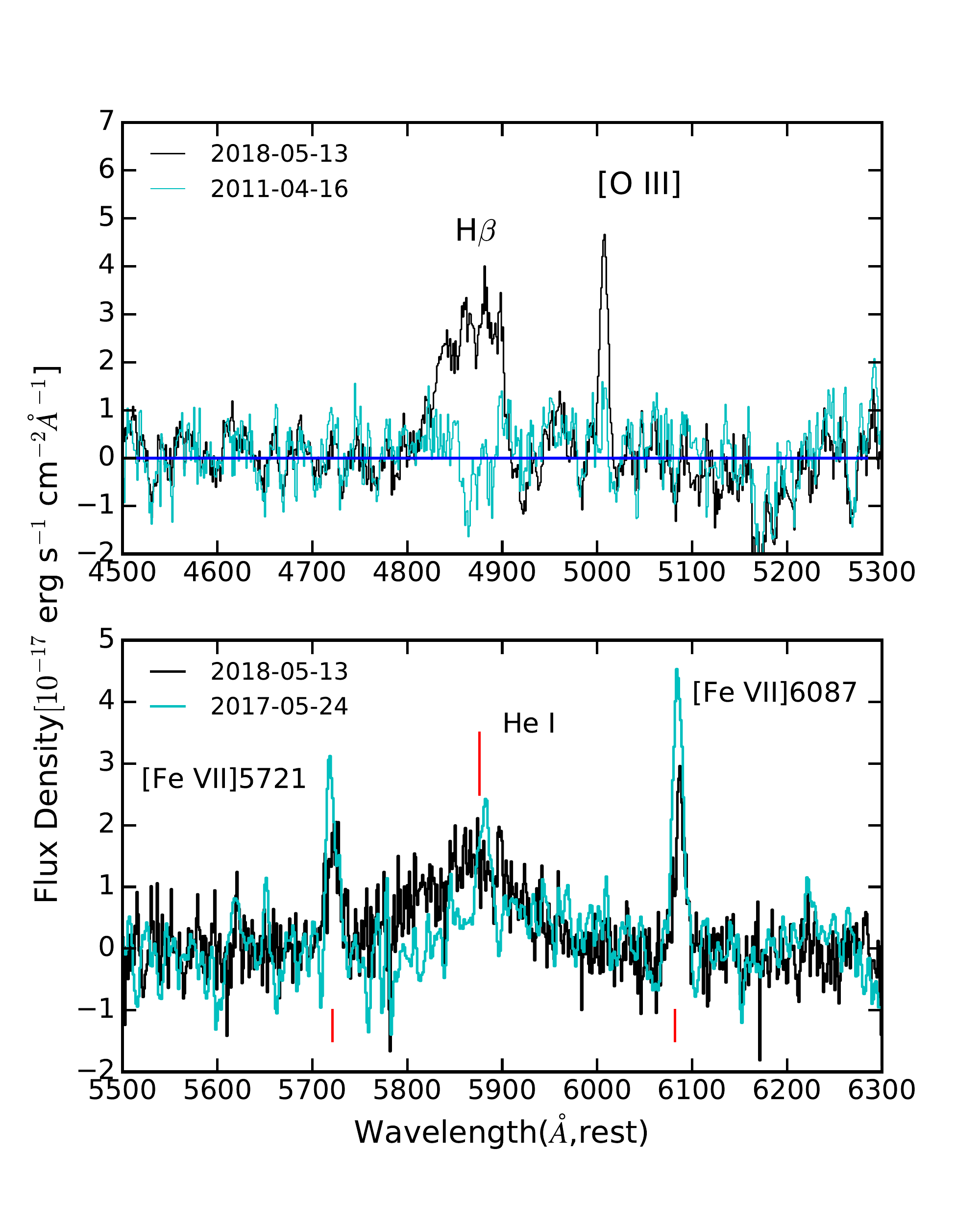}
\caption{ The narrow metal lines, [O\,III]\,5007\,\AA\ and coronal [Fe\,VII], are displayed at the pre-flare (UT 2011-04-16, cyan) and the post-flare (UT 2018-05-13, black) phase. This comparison directly illustrates the temporal change.
\label{fig:narrow}}
\end{figure}

Figure~\ref{fig:narrow} compares the local continuum subtracted spectra around [O\,III]\,5007\,\AA\ doublet and [Fe\,VII]\,5721,6082\,\AA\ lines pre- and post-flare. 
It is clear that after the flare, [O\,III]\,5007\AA\ became stronger, with line flux roughly a factor of 2 higher than that of the SDSS spectrum from 2011. For DBSP spectra, [O\,III]\,5007\AA\ falls near the D550 dichroic beam splitter, thus no reliable information available. We interpret the line flux increase between these two epochs is due to the increase of the ionization continuum. 
 [Fe\,VII]\,5721\,6087\AA\ lines are detected at four different epochs.  Using the same method, we measured $f_{[Fe\,VII]6087}$\,=\,(5.8, 4.3, 4.1, 3.8)$\times10^{-16}$\,erg\,s$^{-1}$\,cm$^{-2}$ at UT 2017-05-24, 2018-01-14, 2018-04-11, and 2018-05-13 respectively.  We conclude that coronal line flux is decreasing with time. The [Fe\,VII] line flux changes in these four epochs echo the variation of the ionization continuum with time (optical LC) and show time delays in relation to the continuum changes because ionizing photons need time to travel the size of the narrow line regions. For details, see Section 4.2.3. As noted in many studies of local AGNs displaying temporal spectral variabilities ({\it e.g.} NGC548 and NGC4151) \citep{Landt2015, Rose2015, Oknyanskij2019}, [Fe\,X] variability is usually associated with that of [Fe\,VII]. However, [Fe\,X] line is not detected in our data. Detailed discussion on various coronal lines are in \S~\ref{sec:coronal}.
  
\subsubsection{Sizes, gas density and temperature of the broad and narrow line emitting regions}
Regardless what physical processes are producing the flare,  several properties of the broad and narrow line emitting regions can be inferred from time lags between continuum and spectral lines, and line ratio diagnostics.

The energetic UV/optical flare is likely to drive kinematic motions. 
The broad Balmer emission lines probably come from photo-ionized gas clumps moving at a velocity of $3000 - 4000$\,km\,s$^{-1}$.  
We can estimate the size of broad line region (BLR) using $\Delta t \times v_{outflow}$, here $\Delta t$ is time and $v_{outflow}$ is cloud velocity.  If we take $\Delta t \leq 300$\,days, the time between the rise time of the optical flare and the first detection of H$\alpha$ (LAMOST data), and assuming clouds moving at $4000$\,km\,s$^{-1}$, we have a BLR radius of $1\times10^{16}$\,cm $\sim 0.003$\,pc.

The size of the narrow line emitting region (NLR) can be estimated in two ways, using time lag and kinematics. The [O\,III] and [Fe\,VII] temporal behaviors can be understood in the context of the time variation of the ionizing continuum.  The post-flare [O\,III]\,5007\,\AA\ line flux is higher than pre-flare because more ionized O$^{+4}$ ions are produced by the UV/optical flare photons. For a NLR with a radius of $R_{N}$, the UV ionizing photons will take $\Delta t$\,=\,$R_{N}/c$ to travel from the center to the line emitting ions. Therefore, there is a time delay of $\Delta t$ between the productions of UV ionizing photons and narrow emission lines. 
The [Fe\,VII] line fluxes are stronger at UT 2017-05-24 than UT 2018-05-13. This is due to higher ionizing UV continuum at the LC peak, between May 2015 and May 2016, prior to the $V$-band LC plateau. The UV ionizing continuum responsible for the 2018-05-13 [Fe\,VII]  emission is likely from the time when the LC has faded, reaching to the plateau in 2017.
Our data can only give an approximate limit on $\Delta t$\,$<$\,770\,days. The narrow region has a size $R_N$\,$<$\,$2\times10^{18}$\,cm $\sim 0.6$\,pc. 

Another way to estimate the NLR size is to use the measured [Fe\,VII] line width. From the Keck spectrum on 2018-05-13, the [Fe\,VII]\,6087\AA\ FWHM is $\sim15.5$\,\AA\ (rest-frame),  which translates to a velocity of $\sim770$\,km\,s$^{-1}$.  Taking into account of the LRIS-R spectral resolution $\Delta \lambda$\,=\,6.9\AA, we have the intrinsic line velocity width $v$\,=\,703\,km\,s$^{-1}$. 
Assuming this line width is due to narrow line emitting gas moving in the gravitational potential of a $2\times10^7M_\odot$ blackhole, we have $R$\,=\,$\frac{M G}{v^2}$\,=\,$5\times10^{17}$\,cm.  These two crude estimates are roughly consistent, and the NLR is more than an order of magnitude larger than the BLR.

\begin{figure}[!ht]
\includegraphics[width=1.05\linewidth]{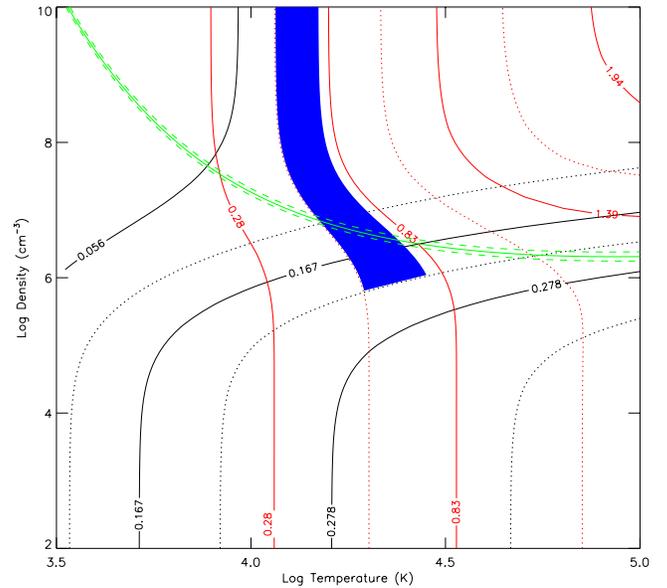}
\caption{ The computed density and temperature of the model ISM with the constraints from the forbidden line ratios. \label{fig:ratios}}
\end{figure}

Regardless of the physical nature of the flare, the observed line ratios from [Fe\,VII] \&\ [O\,III]
can be used to constrain the gas density and temperature, as described in standard text books ({\it e.g} \citet{Osterbrock2006}). These line ratios are computed for a grid of density and temperature. Figure~\ref{fig:ratios} shows   
the [Fe\,VII]\,3760/[Fe\,VII]\,6087 ratio in red lines (dashed line for the averaged value between the two solid lines),  [Fe\,VII]\,5160/[Fe\,VII]\,6087  in black, and [O\,III]\,4959+5007/[O\,III]\,4636 in green.  The solid blue area marks the region which is consistent with the observed coronal line ratios of ${\it f_{\rm [Fe\,VII]3760}/f_{\rm [Fe\,VII]6087}}$ = $0.67^{+0.15}_{-0.13}$, ${\it f_{\rm [Fe\,VII]5160}/f_{\rm [Fe\,VII]6087}}$ $<$ 0.22 (${\it f_{\rm [Fe\,VII]5160}(3\sigma)}$ $<$ 0.8$\times10^{-16}$\,erg\,s$^{-1}$\,cm$^{-2}$), and ${\it f_{\rm [O\,III]4959+5007}/f_{\rm [O\,III]4636}}$ = $2.5^{+2.7}_{-2.3}$.  The $n_e$ and $T_e$ values are constraint by the intersection between the green line and the solid blue area, with $n_e \sim (1.0 - 5)\times10^6$\,cm$^{-3}$ and $T_e \sim (1.5 - 3)\times10^{4}$\,K. 
The derived $T_e$ values suggest that the narrow line emitting gas is photo-ionized by the UV/optical flare.   

Noteworthy is that these density and temperature values are consistent with what have been derived for  coronal line AGNs \citep{Rose2015} as well as post-shocked CSM in type IIn SN, {\it e.g.} SN\,2010jl and SN\,2006jd \citep{Fransson2014,Stritzinger2012}. We discuss this ambiguity further in the section below.

\begin{figure*}[!ht]
\center
\includegraphics[width=1.05\linewidth]{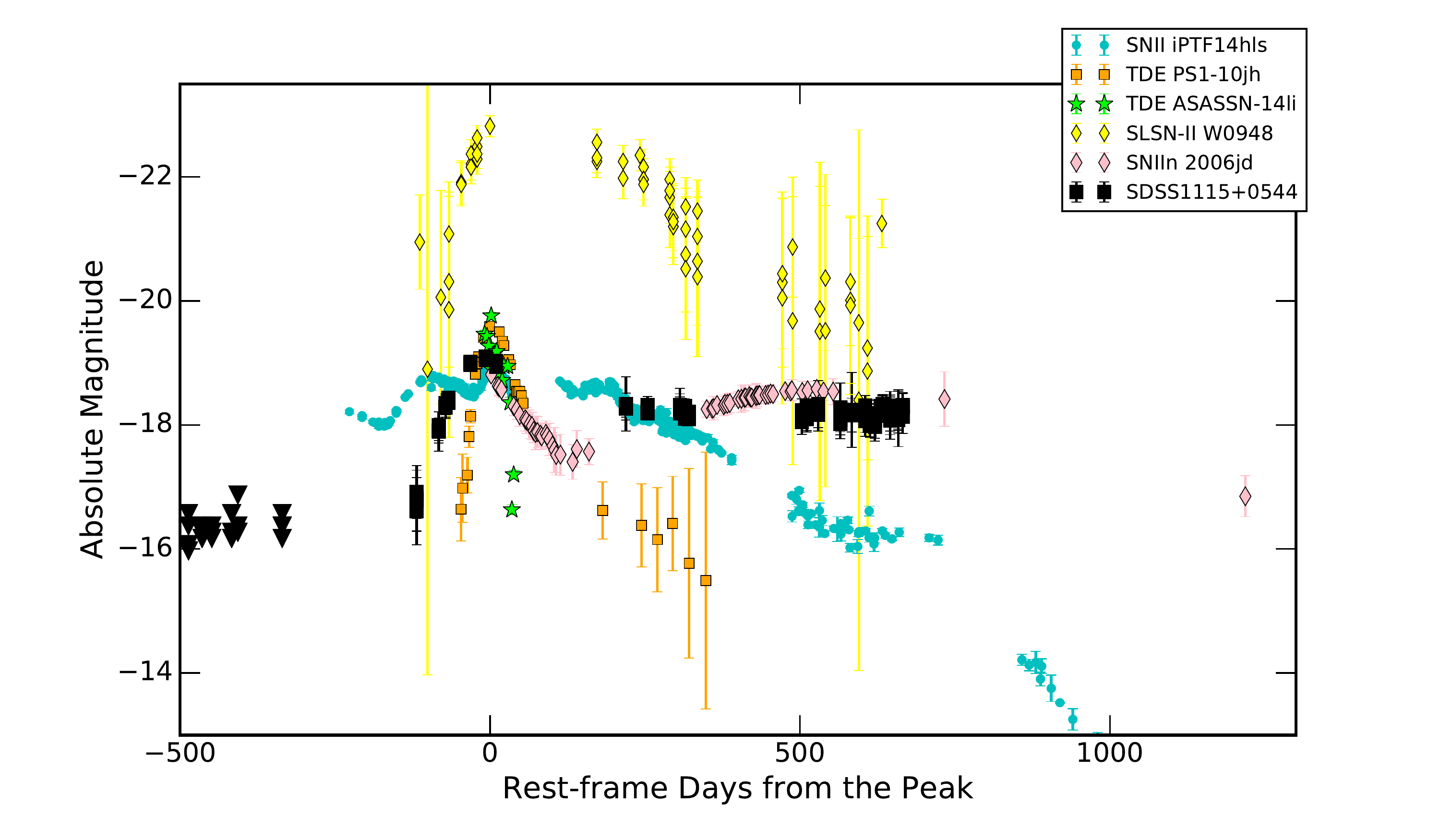}
\caption{The comparison of the SDSS1115+0544 optical LCs with that of SNe and TDEs.  The supernovae are iPTF14hls \citep{Arcavi2017}, SNIIn SN2006jd \citep{Stritzinger2012}, and SLSN-II W0948+0318 \citep{Assef2018}. The two archetypal UV/optical discovered TDEs are ASASSN-14li \citep{Brown2017} and PS1-10jh \citep{Gezari2012}.  The LCs for iPTF14hls, SN2006jd, and PS1-10jh are $r$-band, and SDSS1115+0544, ASASSN-14li and W0948+0318 are in $V$-band.
\label{fig:complc}}
\end{figure*}

\begin{figure}[!ht]
\center
\includegraphics[width=1.05\linewidth]{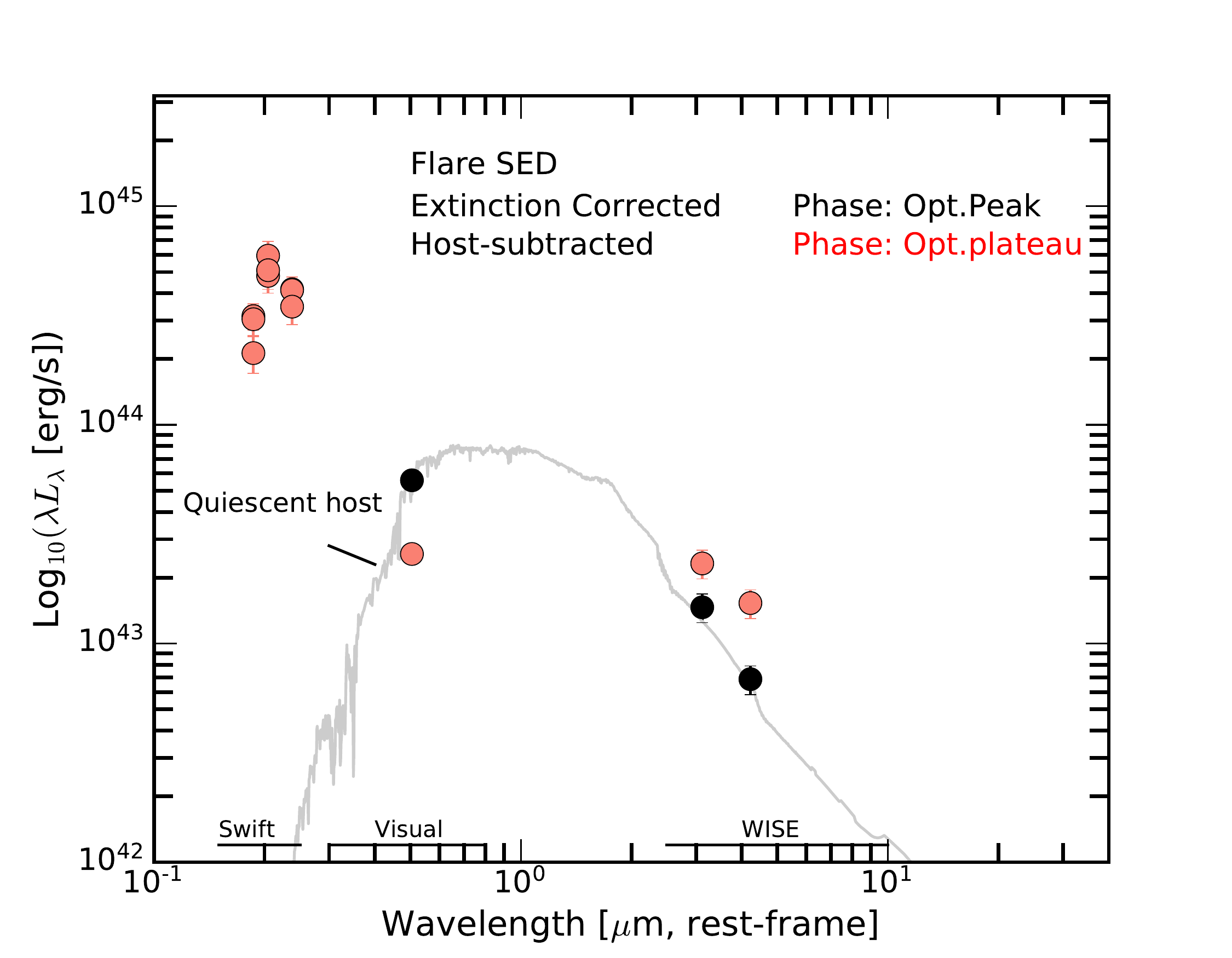}
\caption{The full spectral energy distribution (SED) of the flare from UV to infrared wavelength. We apply the dust extinction corrections to the observed fluxes. The pink and black colors indicate the mid-IR peak and optical peak phases. All photometry are host-subtracted. The {\it Swift} UV photometry is for the flare only because its host has no pre-flare UV emission.  \label{fig:sed2}}
\end{figure}

\subsection{The flare SED from UV to mid-IR and its energetics}
\label{sec:flaresed}

The observed $V$-band peak luminosity is $-19.1$\,mag ($\nu L_{\nu}(V)$\,=\,$1.3\times10^{43}$\,erg\,s$^{-1}$). With $A_V=1.8$, the extinction corrected optical peak luminosity is $-20.9$\,magnitude (AB).  This value is comparable to that of most luminous SN IIn and TDEs (see Figure~\ref{fig:complc}).  The host subtracted, peak mid-IR luminosity is $\sim -22$\,mag (AB) at $3.4$ and $4.6\mu$m (Figure~\ref{fig:lc}), very energetic and consistent with having large dust extinction.  This is much higher than the peak infrared luminosities of many normal, nearby supernovae ($-12$ to $-19$, AB), compiled by SPitzer InfraRed Intensive Transients Survey (SPIRITS) \citep{Kasliwal2017,Tinyanont2016,Fox2016}.  However, two mid-IR luminous supernovae indeed exist. They are SDWFS-MT-1, a self-obscured supernova at $z=0.19$ with $M_{[4.5]} \sim -24.2$ \citep{Koz2010} and SN2003ma, an unusual SN\,IIn in a dense CSM \citep{rest2011}.  Thus, the optical and mid-IR peak luminosities alone can not discriminate the three possible physical models between SN IIn, TDE and ``turn-on'' AGN.

The {\it Swift} UV data were taken after the optical peak date of UT 2015-05-28 (MJD\,=\,57071) and at UT 2017-07-02, 2017-07-13, and 2017-12-25. The UV fluxes do not show significant time variation (see Table~\ref{tab:swiftphot}). 
The {\bf extinction corrected} flare SED is shown in Figure~\ref{fig:sed2} as $\lambda L_\lambda$ from UV, optical to mid-IR for two different phases. The pink symbols correspond to the mid-IR peak phase while the optical LC is at the plateau (Figure~\ref{fig:lc}), and the black symbols are for the earlier, optical peak phase, when the mid-IR LCs are just starting to rise.  

The striking feature of the flare SED is its steep rising SED out to the UV wavelength.  Clearly the bulk of the intrinsic energy from the flare event is coming out at the UV.  Although the observed UV luminosity integrated from 1928\AA\ (UVW2) to 2600\AA\ (UVW1) is only $1.2\times10^{42}$\,erg\,s$^{-1}$, the extinction corrected $L_{UV}$ is high, $1\times10^{44}$\,erg\,s$^{-1}$. Integrated over the flare SED, we infer the extinction corrected, pseudo-bolometric luminosity from the UV to infrared is $\sim$\,$4\times10^{44}$\,erg\,s$^{-1}$, comparable to that of most luminous SN IIn and TDEs.

\section{Discussion:
How to distinquish a SN IIn,  a TDE and a ``turn-on'' AGN?}

Possible physical explanation for the SDSS1115+0544 flare is one of the three scenarios -- an unusual type II SN, a TDE or a ``turn-on'' AGN from a quiescent state.  Identification of the true nature of such a flare can be ambiguous if the available observations are not sufficient.  Here we make extensive comparative studies and highlight the observed properties which have and do not have discriminative powers.

\subsection{Light Curve comparisons with TDE and type II SN}
\label{sec:compare}

Optical light curves of most normal SNe have rise time scales (from explosion to peak) roughly $\sim 10 - 100$\,days, with Superluminous Supernovae (SLSN) being the slowest \citep{Yan2015, Yan2017, DeCia2018}. Normal SNe fade away over time scales of $100 - 300$\,days.  SDSS1115+0544 clearly evolves on much longer time scales, with a bright \&\ 600\,day plateau. 
SN\,II\,P have post-peak plateaus, but only lasting 100\,days or less, then fade rapidly \citep{Rubin2016}.  
SDSS1115+0544 is clearly not a normal type II SN.

However, there are unusual type II (or IIn) SNe, which evolve slowly and their ejecta interact with dense circumstellar medium (CSM). Some examples are iPTF14hls, a type II core collapse SN  \citep{Arcavi2017}, 
SN\,2006jd, an extremely well studied SN IIn \citep{Stritzinger2012} and WISEA\,J094806.56+031801.7 (W0948+0318), a possible H-rich SLSN-II in a mid-IR selected AGN \citep{Assef2018}.  Figure~\ref{fig:complc} compares their LCs with that of SDSS1115+0544 over a time span of 1500\,days.  
Noteworthy is that the post-peak LC of SN\,2006jd (pink diamond) has the most striking similarity to that of SDSS1115+0544 (black square).
Comparisons with the LCs of TDEs are also displayed in Figure~\ref{fig:complc}, including two archetypal UV/optical discovered TDEs, ASASSN-14li and PS1-10jh \citep{Brown2017,Gezari2012}.  

From a pure energetics point of view,  the peak luminosity and total radiated energy from the SDSS1115+0544 flare have little discriminative power among the three possible physical models, and are all consistent with what have been previously observed. The strongest argument against SDSS1115+0544 being a TDE is its long time scales,  for both light curve and spectral line evolution.  Slow rise-time of 120\,days and light curve plateau lasting over 600\,days have not been seen among any confirmed TDEs so far.  However, some TDE light curves seem to show flattening, but only in UV bands, {\it e.g.} AT2018zr (PS16kh), ASASSN-14li \citep{Holoien2018,Sjoert2018, Brown2017}. And their time scales are a factor of 5 or 10 shorter.  The constant H$\alpha$ emission over one year is also unlike any known TDEs.
Larger and more complete TDE samples from future transient surveys, such as ZTF, may find slow evolving TDEs. 
We conclude that the SDSS1115+0544 is probably not a typical TDE and its slow evolving LC by itself cannot rule out the possibility of being an unusual type II SN.

In the context of SN IIn versus TDE, we note relevant points. One is that the empirical characteristics of TDEs are not yet completely known. For example, not well studied flare events include partial destruction of a star or stripping of stars \citep{Campana2015}, and events occurred in AGN host galaxies where larger accretion disks already exists prior to the disruption events \citep{Ivanov2018, Blanchard2017}.  On the nuclear supernovae, they do explode near the centers of galaxies. 
\citet{Villarroel2017} examined the large sample of SNe discovered by PTF since 2009, and found 16 core-collapse supernovae within the projected distances $<10$\,kpc from the centers of the host type-2 AGNs at $z<0.2$. Their results suggest that we can not rule out SDSS1115+0544 being a supernova based on the physical location.

\subsection{Broad Balmer and He\,I lines}

Broad Balmer emission lines with velocity widths of $3000 - 4000$\,km\,s$^{-1}$ have been observed among type II SNe, AGNs and TDEs.  Broad He\,I\,5876\,\AA\  in the SDSS1115+0544 spectrum at +944\,days  was also seen in type IIn SN\,2010jl at +461\,days, and also in SN\,2006jd at +1544\,days \citep{Fransson2014,Stritzinger2012}, as well as in TDE iPTF16fnl \citep{Nadia2017}. 
Therefore, these lines and their velocity widths can not discriminate
the possible three scenarios.  However, the absence of variable narrow H$\alpha$ line is the evidence against the type IIn SN hypothesis because it is the hallmark of the slow moving, dense CSM. However, like the LC morphology, this does not rule out that SDSS1115+0544 may be an unusual type II SN, like iPTF14hls and SN2003ma, where the narrow H$\alpha$ line was absent or emerged only at $>+1000$\,day post-peak \citep{Sollerman2018,rest2011}.

\begin{figure*}[!ht]
\includegraphics[width=1.1\linewidth]{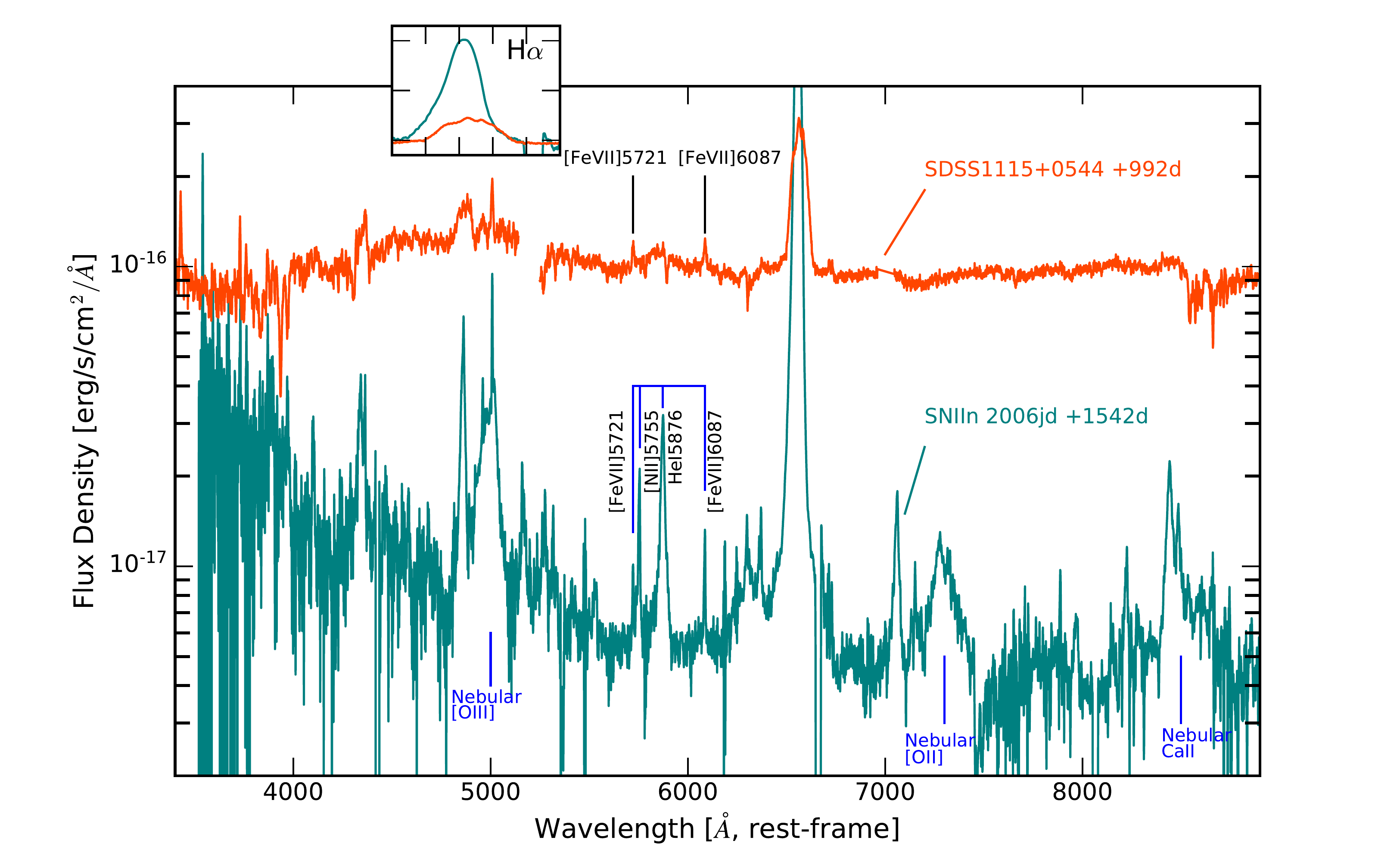}
\caption{The comparison between the SDSS1115+0544 spectrum at +992\,days post-peak and the spectrum at +1542\,days for SNIIn SN2006jd \citep{Stritzinger2012}. \label{fig:compspec}}
\end{figure*}

\subsection{Coronal Fe lines}
\label{sec:coronal}

The formation of coronal [Fe\,VII] lines requires energetic ionizing photons with $E$\,$>$\,100\,eV and a sufficient amount of gas phase Fe atoms which are not locked up by dust grains \footnote{In most galactic ISM, 99\%\ of Fe element is in solid phase due to dust depletion \citep{Savage1996}.}.  Independent of flare models, the detections of [Fe\,VII] lines suggest that SDSS1115+0544 must have had a soft X-ray outburst around the time of the $V$-band flare and the coronal line emitting regions must be relatively dust free during the period between the onset of the flare in 2015 and the last optical spectrum in May 2018.  The dust sublimation radius, the size computed from the mid-IR LC time lag, and the NLR size, are consistent, $\sim$\,$5\times10^{17}$\,cm.  This supports the story that coronal lines are from the inner, ionized gas.

The potential association between coronal lines and TDEs has been extensively studied by  \citet{Komossa2008, Komossa2009, Wang2011, Wang2012, Yang2013}. These studies found several so called ``Extreme Coronal Line Emitters'' (ECLE), which are believed to be possible TDE candidates. Unusually strong [Fe\,X], [Fe\,XI], [Fe\,XIV] and [Fe\,VII] emission lines are found in their initial SDSS spectra, but disappeared in the follow-up spectra taken $4- 9$\,years later. Some of these ECLEs also have variable broad H and He\,II emission.  The time scales of optical LC and spectral evolution for these ECLEs are not constraint because of poor cadence in their spectral observations. 
In SDSS1115+0544, only [Fe\,VII]\,5721,6087\AA\ were observed. [Fe\,X] would have been detected if it had the [Fe\,X]/[Fe\,VII] ratio $\geq 1$ \citep{Wang2012}.  

Coronal lines have been observed for decades among type I and II AGNs \citep{Appenzeller1988, Korista1989, Appenzeller1991,Ferguson1997,Tran2000,Nagao2000,Gelbord2009,Landt2015, Oknyanskij2019}.  The inner wall of the AGN dust torus  is usually heavily irradiated, and just lie at the sublimation radius of the most refractive dust \citep{Krolik1995}. From certain viewing angles, this photo-ionized gas can produce spectra with many coronal lines \citep{Rose2015,Glidden2016}.  The [Fe\,VII] line luminosity from SDSS1115+0544 is comparable to that of Seyfert galaxies \citep{Nagao2000, Gelbord2009}, but its [Fe\,VII]/[O\,III] ratio of 0.4 is much higher than AGNs. This is similar to ECLEs, with unusually strong coronal lines relative to [O\,III]\,5007\AA\ doublet.  

Finally, coronal lines have also been detected in type IIn SNe, especially ones with dense CSM and whose progenitor stars have experienced significant mass losses. 
Figure~\ref{fig:compspec} compares the SDSS1115+0544 spectrum at +992\,days with the late-time spectrum of SN\,IIn 2006jd \citep{Stritzinger2012}.   Coronal Fe lines for SN2006jd only appear at several 100\,days post-peak and become steadily the stronger (higher equivalent width) with time, particularly at +1544\,days. This is mostly due to the declining of the continuum.  Similar to AGNs, photoionization by X-rays from shocks is responsible for producing coronal emitting gas.  

In conclusion, we find that high ionization coronal lines have no discriminative power among SN IIn, TDE and AGNs.

\subsection{Evidence for ``turn-on'' AGN -- constant UV emission and lack of O \&\ Ca broad lines at late-times}

\begin{figure}[!ht]
\center
\includegraphics[width=1.05\linewidth]{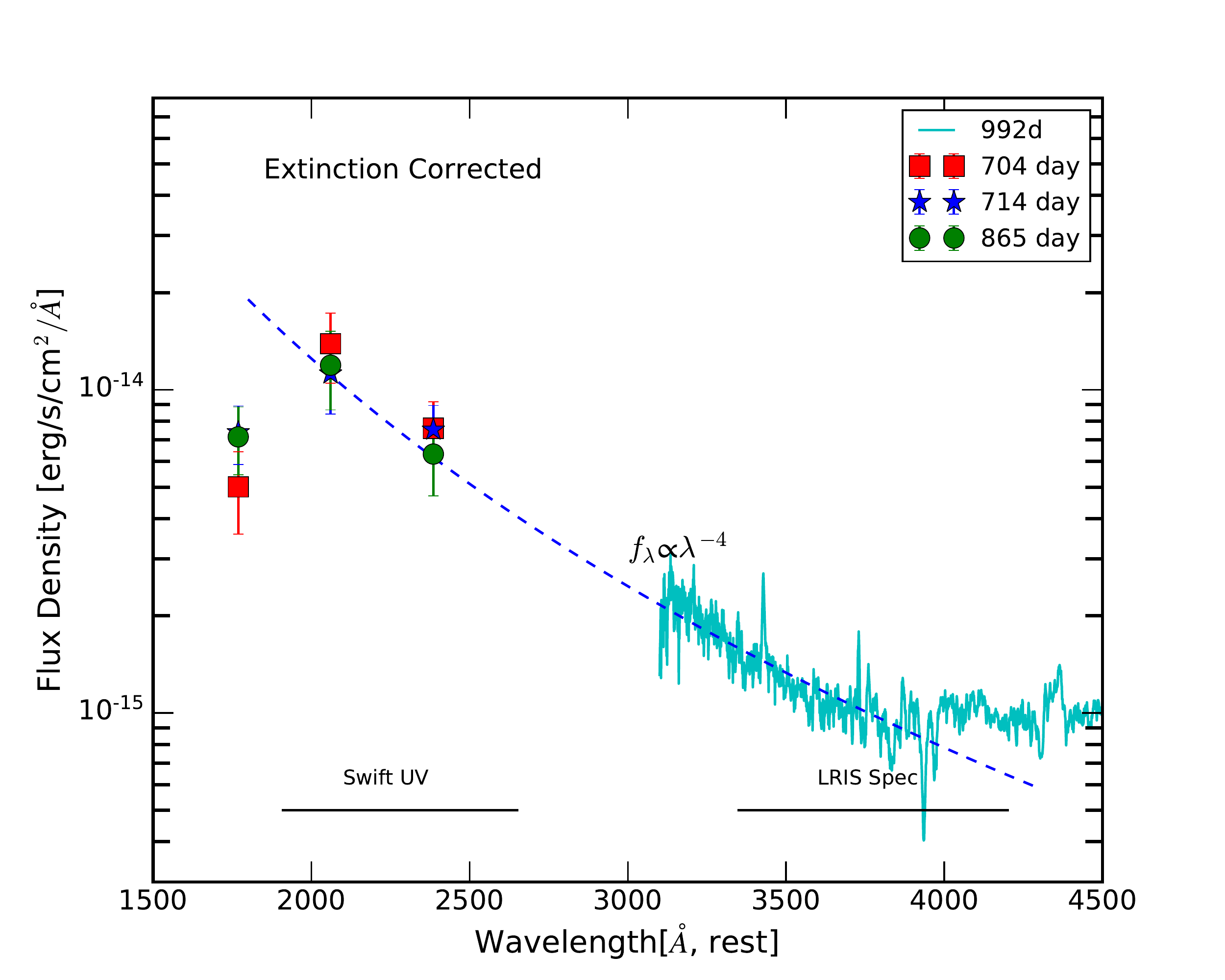}
\caption{The dust extinction corrected UV SED of the flare measured by {\it Swift} photometry at three epochs ($700 - 865$\,days from the peak) and the Keck LRIS blue side of the spectrum at 992\,days from the peak.  The {\it Swift} UV fluxes do not show variation between epochs.  Assuming no temporal variation, the spectral slope between 1900 - 4300\AA\ is shown in blue dashed line, with $f_{\lambda} \propto \lambda^{-3}$.
 \label{fig:uvsed}}
\end{figure}

The strongest evidence for a ``turn-on'' AGN model and against a type IIn SN or a TDE model is the strong, constant UV continuum emission at very late times. The host galaxy has no pre-flare UV emission, and the newly formed UV emission is associated with the flare event.  Figure~\ref{fig:uvsed}  shows the zoom-in portion of the intrinsic (dust extinction corrected) UV SED for the flare at the post-peak $+700$ to $+992$\,days.  The three {\it Swift} UV fluxes show little changes with time, and the UV spectral slope is extremely steep, with $f_{\lambda} \propto \lambda^{-4}$.   UV emissions from supernova and TDE can be strong at early times and usually fade quickly with time.  This observation supports the AGN scenario because the newly ``turn-on'' accretion disk can be an efficient source of UV emission. 

One additional argument against SN IIn model is lack of broad emission lines from O and Ca elements in
the +992\,day Keck spectrum (Figure~\ref{fig:compspec}).  Late-time variable O and Ca line emission are very common
among SN IIn. The examples of slow evolving, unusual type II SNe, such as 
SN\,2006jd, iPTF14hls, SN2010jl, SN2003ma and SN2005ip, 
all display changes in the continua and spectral features, with broad emission lines from H, O and Ca elements emerging at late phases. 

In summary, the flare in SDSS1115+0544 is likely a ``turn-on'' AGN, changing from a quiescent galaxy to a type 1 AGN.

\subsection{SDSS1115+0544 as a ``turn-on'' AGN: accretion rate change and space density}

In this scenario, how much increase would be the blackhole accretion rate?
Taking into account of dust extinction corrections to both optical and {\it Swift} UV fluxes, the bolometric luminosity for the flare $L_{bol}(flare)$ is $\sim$\,$10^{44}$\,erg\,s$^{-1}$.  If this flare is due to matter being accreted onto the blackhole, we have $L_{bol}(flare)$\,=\,$\eta \Delta \dot{m} c^2$ and the matter to radiative efficiency $\eta \sim 0.1$. This gives the change in accretion rate $\Delta \dot{m}$\,$\sim0.017M_\odot$/yr.  This is a reasonable value for AGNs.  

In the standard picture of broad line regions around a blackhole, H$\beta$ line width is correlated with the central blackhole mass. We adopt the following relation established by \citet{Vestergaard2006}, where
$\rm  log_{10} M_{BH}(H\beta) =  
log_{10} \Bigg\{ \Big[\frac{FWHM(H\beta)}{1000\,km\,s^{-1}} \Big]^2 \Big[\frac{L(H\beta)}{10^{42}\,ergs\,s^{-1}}\Big]^{0.63} \Bigg\} \\ 
+ (6.67\pm0.03).$
With $\rm FWHM(H\beta)$\,=\,3271\,km\,s$^{-1}$ and the extinction corrected L$(H\beta)$\,=\, $3.5\times10^{41}$\,erg\,s$^{-1}$ ($\rm A_{H\beta}$\,=\,2.1\,mag),  the inferred $\rm M_{BH}$ is  $2.6\times10^7M_\odot$, confirming the value based on the SDSS velocity dispersion. Using $\rm M_{BH}$ and $L_{bol}$ values, we estimate the Eddington ratio $\lambda_{Edd} = L_{bol}/L_{Edd} = 0.03$, indicating low accretion rate.


If SDSS1115+0544 is a ``turn-on'' AGN, how are its H$\alpha$ and [O\,III] line luminosities compared with that of known AGNs and QSOs? For low $z\sim0.1$ type 1 AGN,  its typical $L_{5100\AA}$ to $L_{H\alpha}$ ratio is $\sim 19$ at $L_{5100\AA}\sim10^{44}$\,erg\,s$^{-1}$, based on the correlation derived by 
\citet{Green2005}, $L_{H\alpha}=(5.25\pm0.02)\times10^{42} (\frac{L_{5100}}{10^{44}})^{1.157}$\,erg\,s$^{-1}$, 
and the typical $L_{5100\AA}$ to $L_{[OIII]}$ ratio is 320 \citep{Kauffmann2005}.  
At the plateau phase, SDSS1115+0544 flare has a $V$-band luminosity (host-subtracted) of $\sim1\times10^{43}$\,erg\,s$^{-1}$, $L_{5100\AA}$/$L_{H\alpha}$ of $\sim 23$ and $L_{5100\AA}$/$L_{[OIII]}$\,$\sim$\,588.  This implies that SDSS1115+0544 flare has a typical $L_{H\alpha}$ as that of type 1 AGN, but its $L_{[O\,III]}$ is too weak, $1.7\times10^{40}$\,erg\,s$^{-1}$, just below the $10^7L_\odot$ dividing line between weak and strong AGNs, as defined by SDSS AGN studies \citep{Kauffmann2005}.  Thus we conclude that similar to iPTF16bco, the NLR for SDSS1115+0544 is still developing, consistent with brightening of [O\,III] strength. Future optical spectra will further confirm this result.

In addition, SDSS1115+0544 optical LC suggests that its continuum has changed from quiescent to the peak of flaring in only 200\,days. For a radiation-dominated, Shakura-Sunyaev disk, the viscous radial inflow time scale \citep{Shakura1973,LaMassa2015}, 

\begin{equation}
t_{infl} = 978\,yr [\frac{\alpha}{0.1}]^{-1}[\frac{\lambda_{Edd}}{0.03}]^{-2}[\frac{\eta}{0.1}]^{2}[\frac{r}{50r_g}]^{7/2}[\frac{M_{BH,7}}{2.0}]
\end{equation}

where $\alpha$ is the Shakura \&\ Sunyaev viscosity parameter, $r_g$ is
the gravitational radius ($GM/c^2$).  If $\alpha$, $\eta$ and $r$ are assumed to be 0.1, 0.1 and 50$r_g$ (optical emission typically comes from $50-100r_g$) respectively, the viscous inflow time scale at $50r_g$ is 978\,years, much longer than 200\,days continuum ``turn-on'' time scale measured from the SDSS1115+0544 LC. This ``viscosity crisis'' has been noted, and is the focus of a recent paper by \citet{Lawrence2018}.

SDSS1115+0544 is a case of a quiescent LINER transitioning into an AGN in a short time scale $<1$\,year.  It is the first rapid ``turn-on'' AGN with a well constraint rise time scale based on its extensive CRTS light curve. 
iPTF16bco is another such an event at $z=0.237$ \citep{Gezari2017}. Its quiescent host is also a LINER, and the transformation was characterized by optical flare (iPTF LC), increasing UV continuum emission and newly formed broad H$\alpha$ emission, similar to that of SDSS1115+0544.  It is attempting to speculate that the short ``turn-on'' time scales found in these two cases may be associated with LINERs where rapid accretion rate changes or disk instability may likely occur.  However, better understanding requires more such events to be discovered by ongoing and future transient surveys.   

AGNs have been known and studied more than a half-century.  The actively accreting phase, {\it i.e.} AGN phase, is thought to last roughly  a few times $10^7$\,years (so called duty cycle) \citep{Combes2000, Haehnelt1993}.  ``Turn-on'' AGNs like SDSS1115+0544 and iPTF16bco have a flare time scale of $200 - 300$\,days.  By scaling these two time scales, we infer that ``turn-on'' AGN is only $10^{-7}$ of the parent AGN population.  Counting low luminosity AGNs, the space density of Seyfert 1 is roughly $10^{-4}$\,Mpc$^{-3}$ at low redshift \citep{Osterbrock2006}. Thus, events like SDSS1115+0544 should have space density of $10^{-11}$\,Mpc$^{-3}$.   
However, without any systematic searches, two rapid ``turn-on'' type 1 AGN/QSO have been found with the flaring peak time separated by only a year. This may indicate that our calculation could significantly under-estimate the true value.  Perhaps the AGN duty cycle is not the proper time scale to use for this calculation.  Because duty cycle of $10^7$\,year only counts for the total time when AGN is active. This does not distinguish various flare episodes which will have a distribution of different time scales.  Upcoming large transient surveys should produce actual measurements of this distribution.

\section{Summary}
\label{sec:discuss}

SDSS1115+0544 is a transient first selected as a part of our systematic search for infrared flare galaxies within the SDSS area.  The host galaxy is a quiescent, early type galaxy at $z=0.0899$ with M$_{stars} \sim 3.5\times10^{10}M_\odot$, M$_{BH} \sim 2\times10^7M_\odot$, SFR$_{24\mu m}$ of $\sim0.2 M_\odot$/yr. Its SDSS optical spectrum has the characteristics of a LINER with a non-accreting blackhole and weak narrow forbidden emission lines.
In Janurary 2015, SDSS1115+0544 underwent a strong optical flare, with a rise of 2.5 magnitude ($V$-band) over $\sim 120$\,days. Declining from the peak magnitude (host-subtracted) of 
After +200\,days post-peak, its optical LC reached a prolonged plateau, 1.8 magnitude brighter than the quiescent level and lasting over 600\,days. This optical flare was followed by mid-infrared flares at $3.4$ and $4.6\mu$m with amplitudes of $\geq 0.5$\,magnitude  over 600\,days. The time lag between the infrared and optical peak is $\sim180$\,days.  SDSS1115+0544 displayed excess UV continuum emission which were measured by {\it Swift} UV observations at three epochs, $>+730$\,days post-peak (optical LC). This excess UV emission shows no temporal variation, likely to be on a plateau similar to the optical. A  series of optical spectra of this flare event were obtained between 2016-01-06 and 2018-05-13. Compared to the SDSS spectrum taken in 2011, the new spectra taken after the optical flare show a rich suite of newly formed emission lines, including broad Balmer and helium lines with velocity widths $\sim3750$\,km\,s$^{-1}$, and narrow high ionization forbidden lines such as [Fe\,VII], [Ne\,V].   
 
We made extensive comparisons of the SDSS1115+0544 properties with known TDEs and unusual type II SN. 
Although the newly formed coronal Fe lines link this event to Extreme Coronal Line Emitters (ECLE), the slow evolving optical LC over 1200\,days and its non-variant broad Balmer lines at very late-times argue against SDSS1115+0544 flare being a typical UV/optically selected TDE. 
We find that LC energetics, morphology, time scale as well as spectral features, including broad Balmer emission and narrow coronal lines, can not serve as clear discriminants between ``turn-on'' AGN and an unusual type II SN. This highlights the ambiguity in identifications of many transients discovered in the nuclear regions of quiescent galaxies.

In the case of SDSS1115+0544, the strongest evidence against the scenario of unusual type II SNe is the strong UV continuum observed at $+700$ to $+900$\,days without any temporal variations.  In addition, our Keck spectrum at $+992$\,days does not show any broad O and Ca emission lines commonly seen in the late-time spectra of type II SN, as in examples of unusual type II SN SN2006jd, iPTF14hls, SN2005ip and SN2010jl. We conclude that the flare in SDSS1115+0544 is from a newly ``turn-on'' accretion disk around the supermassive blackhole, transforming the previously quiescent early type galaxy into a type 1 AGN, with  a sub-Eddington accretion rate of $\sim$0.02$M_\odot$/yr.   

This newly ``turn-on'' accretion disk produces significantly soft X-ray and UV photons which are responsible for both continuum flare, the emergence of broad Balmer emission lines as well as narrow high ionization metal lines.   The radiative feedback from the accretion can drive gas outflows, naturally producing a broad Balmer line emitting region with a size likely $<10^{16}$\,cm.  SDSS1115+0544 contains a substantial dusty ISM, with a column density of $3\times10^{21}$\,cm$^{-2}$ ($\rm A_V = 1.8$). This dusty medium produces the infrared echos of the optical flare, generating the infrared variability. The time lag between the IR and optical LCs implies that the dusty ISM is about $5\times10^{17}$\,cm from the central accretion disk.  

It is interesting to note that if the rise time scale determines the AGN ``turn-on'' time scale, it is very short for SDSS1115+0544, $<200$\,days from the well sampled CRTS LC.  This discovery suggests that the changing state phenomenon among AGNs could be associated with violent instabilities in accretion disks occurring on time scales of a few 100\,days. More events such as SDSS1115+0544 from future transient surveys should provide better constraints on the event rates as well as theoretical understanding of accretion disks around supermassive blackholes.

\acknowledgements

We thank A. Ho, M. Kuhn from Caltech, T. Hung, S. Frederick from University of Maryland, for helping with taking some of the spectral observations.  This research has made use of the NASA/ IPAC Infrared Science Archive, which is operated by the Jet Propulsion Laboratory, California Institute of Technology, under contract with the National Aeronautics and Space Administration.
We also use the data products from the Wide-field
Infrared Survey Explorer, which is a joint project of the
University of California, Los Angeles, and the Jet Propulsion
Laboratory/California Institute of Technology, funded by the
National Aeronautics and Space Administration. This paper
also utilized the publicly available SDSS data sets. Funding
for the SDSS and SDSS-II has been provided by the Alfred
P. Sloan Foundation, the Participating Institutions, the National
Science Foundation, the US Department of Energy, the
National Aeronautics and Space Administration, the Japanese
Monbukagakusho, the Max Planck Society, and the Higher Education
Funding Council for England. The SDSS Web site is
http://www.sdss.org/. The SDSS is managed by the Astrophysical
Research Consortium for the Participating Institutions. The
Participating Institutions are the American Museum of Natural
History, Astrophysical Institute Potsdam, University of Basel,
University of Cambridge, Case Western Reserve University,
University of Chicago, Drexel University, Fermilab, the Institute
for Advanced Study, the Japan Participation Group, Johns
Hopkins University, the Joint Institute for Nuclear Astrophysics,
the Kavli Institute for Particle Astrophysics and Cosmology,
the Korean Scientist Group, the Chinese Academy of Sciences
(LAMOST), Los Alamos National Laboratory, the Max-PlanckInstitute
for Astronomy (MPIA), the Max-Planck-Institute for
Astrophysics (MPA), New Mexico State University, Ohio State
University, University of Pittsburgh, University of Portsmouth,
Princeton University, the United States Naval Observatory, and
the University of Washington.
This research has made use of the NASA/IPAC Extragalactic Database (NED) which is operated by the Jet Propulsion Laboratory, California Institute of Technology, under contract with the National Aeronautics and Space Administration. Some of the data presented herein were obtained at the W.M. Keck Observatory, which is operated as a scientific partnership among the California Institute of Technology, the University of California and the National Aeronautics and Space Administration. The Observatory was made possible by the generous financial support of the W.M. Keck Foundation. The authors wish to recognize and acknowledge the very significant cultural role and reverence that the summit of Mauna Kea has always had within the indigenous Hawaiian community.  We are most fortunate to have the opportunity to conduct observations from this mountain.

{\it Facilities:} \facility{Palomar}, \facility{Keck, WISE}, {\it Swift}

\bibliography{mySLSNref}

\clearpage

\begin{deluxetable*}{cccccccc}
\tabletypesize{\footnotesize}
\tablewidth{0pt}
\tablecaption{The {\it Swift} UV photometry
\label{tab:swiftphot}}
\tablehead{
\colhead{Obs.Date} &
\colhead{MJD} &
\colhead{Filter} &
\colhead{$\nu$} &
\colhead{$f_\nu$} &
\colhead{$\delta f_\nu$} & 
\colhead{AB mag} & 
\colhead{$\rm \delta mag$} \\
 &  days &   & $10^{15}$\,Hz & mJy  & mJy & mag & mag \\
}
\startdata
2017-07-02 & 57936.943 & UVW2 & 1.475 & 0.0084 &   0.0016 & 21.59   &  0.21  \\
2017-07-02 & 57936.933 & UVM2 & 1.345 & 0.0141 &   0.0023 & 21.03   &  0.18  \\
2017-07-02 & 57936.953 & UVW1 & 1.157 & 0.0209 &   0.0029 & 20.58   &  0.15  \\
2017-07-13 & 57947.872 & UVW2 & 1.475 & 0.0124 &   0.0017 & 21.16   &  0.15   \\
2017-07-13 & 57947.875 & UVM2 & 1.345 & 0.0114 &   0.0019 & 21.26   &  0.18   \\
2017-07-13 & 57947.878 & UVW1 & 1.157 & 0.0207 &   0.0026 & 20.60   &  0.14   \\
2017-12-26 & 58112.816 & UVW2 & 1.475 & 0.0120 &   0.0019 & 21.20   &  0.17   \\
2017-12-26 & 58112.817 & UVM2 & 1.345 & 0.0121 &   0.0022 & 21.19   &  0.20   \\
2017-12-26 & 58112.819 & UVW1 & 1.157 & 0.0174 &   0.0030 & 20.79   &  0.18   \\

\enddata
\end{deluxetable*}

\begin{deluxetable*}{cccccc}
\tabletypesize{\footnotesize}
\tablewidth{0pt}
\tablecaption{The Spectroscopic Observation Log
\label{tab:speclog}}
\tablehead{
\colhead{Obs.Date} &
\colhead{MJD} &
\colhead{Phase$^a$} &
\colhead{Instrument} &
\colhead{$\Delta \lambda$} &
\colhead{Inst. Res.$^b$} \\
 &  days & days &  & \AA &  \\
}
\startdata
2011-04-16 & 52326 & -4444 & SDSS   & 3100 - 9000   & 1800 \\
2016-01-06 & 57393 & 204.6 & LAMOST & 3100 - 9000   & 2500 \\
2017-05-24 & 57898 & 668.0 & DBSP   & 3100 - 9500   & 700/1200  \\
2017-11-11 & 58068 & 823.9 & DBSP   & 3100 - 9500   & 700/1200 \\
2017-11-25 & 58081 & 835.9 & DBSP   & 3400 - 9500   & 700/1200 \\
2018-01-14 & 58132 & 882.7 & DBSP   & 3400 - 9500   & 700/1200 \\
2018-04-11 & 58218 & 961.6 & DBSP   & 3100 - 9500   & 700/1200  \\
2018-05-13 & 58251 & 992.3 & LRIS   & 3100 - 10000  & 1000/1000 \\

\tablenotetext{a}{The rest-frame phases is relative to the peak dates of MJD = 57170.}
\tablenotetext{b}{Instrument resolution is shown for blue and red side, computed as $\lambda/\delta \lambda$ at 4000 \&\ 7000\AA.}
\enddata
\end{deluxetable*}

\begin{deluxetable*}{cccc}
\tabletypesize{\footnotesize}
\tablewidth{0pt}
\tablecaption{The Emission Line Measurements$^a$
\label{tab:metaline}}
\tablehead{
\colhead{Line Name} &
\colhead{Wavelength} &
\colhead{Line Flux} &
\colhead{FWHM}  \\
 &  \AA & erg\,s$^{-1}$\,cm$^{-2}$ & \AA  \\
}
\startdata
H$\alpha$   &  6563  & $(1.7\pm0.007)$e-14 & 85 \\
H$\beta$    &  4861  & $(2.8\pm0.06)$e-15  & 78 \\
H$\gamma$   &  4341  & $(5.9\pm0.6)$e-16  & 53 \\
$He\,I$     & 5876   & $(1.9\pm0.07)$e-15 & 60 \\
$[$Fe\,X$]^a$    & 6373    & $<$9.2e-17  & $<7.9$ \\
$[$Fe\,VII$]$  & 6087    & $(3.6\pm0.3)$e-16  & 16.9 \\
$[$Fe\,VII$]$  & 5721    & $(3.1\pm0.3)$e-16  & 15.2 \\
$[$O\,III$]$   & 5007    & $(6.2\pm0.1)$e-16  & 12.0 \\
$[$O\,III$]$   & 4959    & $(1.9\pm0.1)$e-16  & 14.0 \\
$[$O\,III$]$   & 4364    & $(3.3\pm0.2)$e-16  & 11.5     \\
$[$Ne\,III$]$  & 3869    & $(2.7\pm0.2)$e-16  & 11.4 \\
$[$Fe\,VII$]$  & 3760    & $(2.4\pm0.3)$e-16  & 10.3  \\
$[$Ne\,V$]^b$  & 3425    & $(7.6\pm0.4)$e-16  & 9.0 \\
$[$Ne\,V$]^b$  & 3346    & $(2.8\pm0.3)$e-16  & 7.2 \\
He\,II$^b$     & 3203    & $(2.3\pm0.3)$e-16  & 7.3 \\
O\,III$^b$     & 3133    & $(4.4\pm0.5)$e-16  & 11.9 \\
\hline \hline
$[$S\,II$]$    & 6731    & 5.5e-17 & 6.3 \\
$[$S\,II$]$    & 6717    & 9.4e-17 & 9.8 \\
$[$O\,II$]$    & 3727    & 4.0e-16  & 5.5 \\

\tablenotetext{a}{All of the line measurements are done using the Keck spectrum taken on 2018-05-13.}
\enddata
\end{deluxetable*}

\end{document}